\documentclass[sigconf, screen]{acmart}  %
\usepackage{acmart-taps}

\AtBeginDocument{%
  }

\copyrightyear{2024}
\acmYear{2024}
\setcopyright{cc}  %
\setcctype[4.0]{by}  %
\acmConference[CHI '24]{Proceedings of the CHI Conference on Human Factors in Computing Systems}{May 11--16, 2024}{Honolulu, HI, USA}
\acmBooktitle{Proceedings of the CHI Conference on Human Factors in Computing Systems (CHI '24), May 11--16, 2024, Honolulu, HI, USA}
\acmDOI{10.1145/3613904.3642001}
\acmISBN{979-8-4007-0330-0/24/05}

\begin{document}

\title{DeepSee: Multidimensional Visualizations of Seabed Ecosystems}

\author{Adam Coscia}
\orcid{0000-0002-0429-9295}
\affiliation{%
  \institution{Georgia Institute of Technology}
  \streetaddress{North Avenue}
  \city{Atlanta}
  \state{Georgia}
  \country{USA}
  \postcode{30332}
}%
\email{acoscia6@gatech.edu}

\author{Haley M. Sapers}
\orcid{0000-0002-1797-1722}
\affiliation{%
  \institution{Division of Geological and Planetary Sciences, California Institute of Technology}
  \streetaddress{1200 East California Boulevard}
  \city{Pasadena}
  \state{California}
  \country{USA}
  \postcode{91125}
}%
\email{hsapers@caltech.edu}

\author{Noah Deutsch}
\orcid{0009-0005-8303-834X}
\affiliation{%
  \institution{Harvard University}
  \streetaddress{Massachusetts Hall}
  \city{Cambridge}
  \state{Massachusetts}
  \country{USA}
  \postcode{02138}
}%
\email{ndeutsch@mde.harvard.edu}

\author{Malika Khurana}
\orcid{0009-0008-7560-741X}
\affiliation{%
  \institution{The New York Times Company}
  \streetaddress{620 Eighth Avenue}
  \city{New York}
  \state{New York}
  \country{USA}
  \postcode{10018}
}%
\email{malika.khurana@nytimes.com}

\author{John S. Magyar}
\email{jmagyar@caltech.edu}
\orcid{0000-0002-3586-8286}
\author{Sergio A. Parra}
\email{sparra@caltech.edu}
\orcid{0000-0002-2637-7960}
\author{Daniel R. Utter}
\email{dutter@caltech.edu}
\orcid{0000-0003-3322-7108}
\author{Rebecca L. Wipfler}
\email{rwipfler@caltech.edu}
\orcid{0000-0003-0602-1753}
\affiliation{%
  \institution{Division of Geological and Planetary Sciences, California Institute of Technology}
  \streetaddress{1200 East California Boulevard}
  \city{Pasadena}
  \state{California}
  \country{USA}
  \postcode{91125}
}%

\author{David W. Caress}
\email{caress@mbari.org}
\orcid{0000-0002-6596-9133}
\author{Eric J. Martin}
\email{emartin@mbari.org}
\orcid{0000-0002-2398-4503}
\author{Jennifer B. Paduan}
\email{paje@mbari.org}
\orcid{0000-0002-4242-5432}
\affiliation{%
  \institution{Monterey Bay Aquarium Research Institute}
  \streetaddress{7700 Sandholdt Road}
  \city{Moss Landing}
  \state{California}
  \country{USA}
  \postcode{95039}
}%

\author{Maggie Hendrie}
\orcid{0009-0001-1916-3761}
\affiliation{%
  \institution{ArtCenter College of Design}
  \streetaddress{1700 Lida Street}
  \city{Pasadena}
  \state{California}
  \country{USA}
  \postcode{91103}
}%
\email{maggie.hendrie@artcenter.edu}

\author{Santiago Lombeyda}
\email{santiago@caltech.edu}
\orcid{0000-0003-0684-7889}
\author{Hillary Mushkin}
\email{hmushkin@caltech.edu}
\orcid{0000-0002-1967-1790}
\affiliation{%
  \institution{California Institute of Technology}
  \streetaddress{1200 East California Boulevard}
  \city{Pasadena}
  \state{California}
  \country{USA}
  \postcode{91125}
}%

\author{Alex Endert}
\orcid{0000-0002-6914-610X}
\affiliation{%
  \institution{Georgia Institute of Technology}
  \streetaddress{North Avenue}
  \city{Atlanta}
  \state{Georgia}
  \country{USA}
  \postcode{30332}
}%
\email{endert@gatech.edu}

\author{Scott Davidoff}
\orcid{0000-0002-4417-7268}
\affiliation{%
  \institution{Jet Propulsion Laboratory, California Institute of Technology}
  \streetaddress{4800 Oak Grove Drive}
  \city{Pasadena}
  \state{California}
  \country{USA}
  \postcode{91109}
}%
\email{scott.davidoff@jpl.nasa.gov}

\author{Victoria J. Orphan}
\orcid{0000-0002-5374-6178}
\affiliation{%
  \institution{Division of Geological and Planetary Sciences, California Institute of Technology}
  \streetaddress{1200 East California Boulevard}
  \city{Pasadena}
  \state{California}
  \country{USA}
  \postcode{91125}
}%
\email{vorphan@caltech.edu}

\renewcommand{\shortauthors}{Coscia, et al.}

\begin{abstract}
  Scientists studying deep ocean microbial ecosystems use limited numbers of sediment samples collected from the seafloor to characterize important life-sustaining biogeochemical cycles in the environment.
  Yet conducting fieldwork to sample these extreme remote environments is both expensive and time consuming, requiring tools that enable scientists to explore the sampling history of field sites and predict where taking new samples is likely to maximize scientific return.
  We conducted a collaborative, user-centered design study with a team of scientific researchers to develop \textit{DeepSee}, an interactive data workspace that visualizes 2D and 3D interpolations of biogeochemical and microbial processes in context together with sediment sampling history overlaid on 2D seafloor maps.
  Based on a field deployment and qualitative interviews, we found that \textit{DeepSee} increased the scientific return from limited sample sizes, catalyzed new research workflows, reduced long-term costs of sharing data, and supported teamwork and communication between team members with diverse research goals.
\end{abstract}



\maketitle

\section{Introduction}
\label{sec:introduction}

\begin{figure*}[!t]
  \centering
  \includegraphics[width=\linewidth]{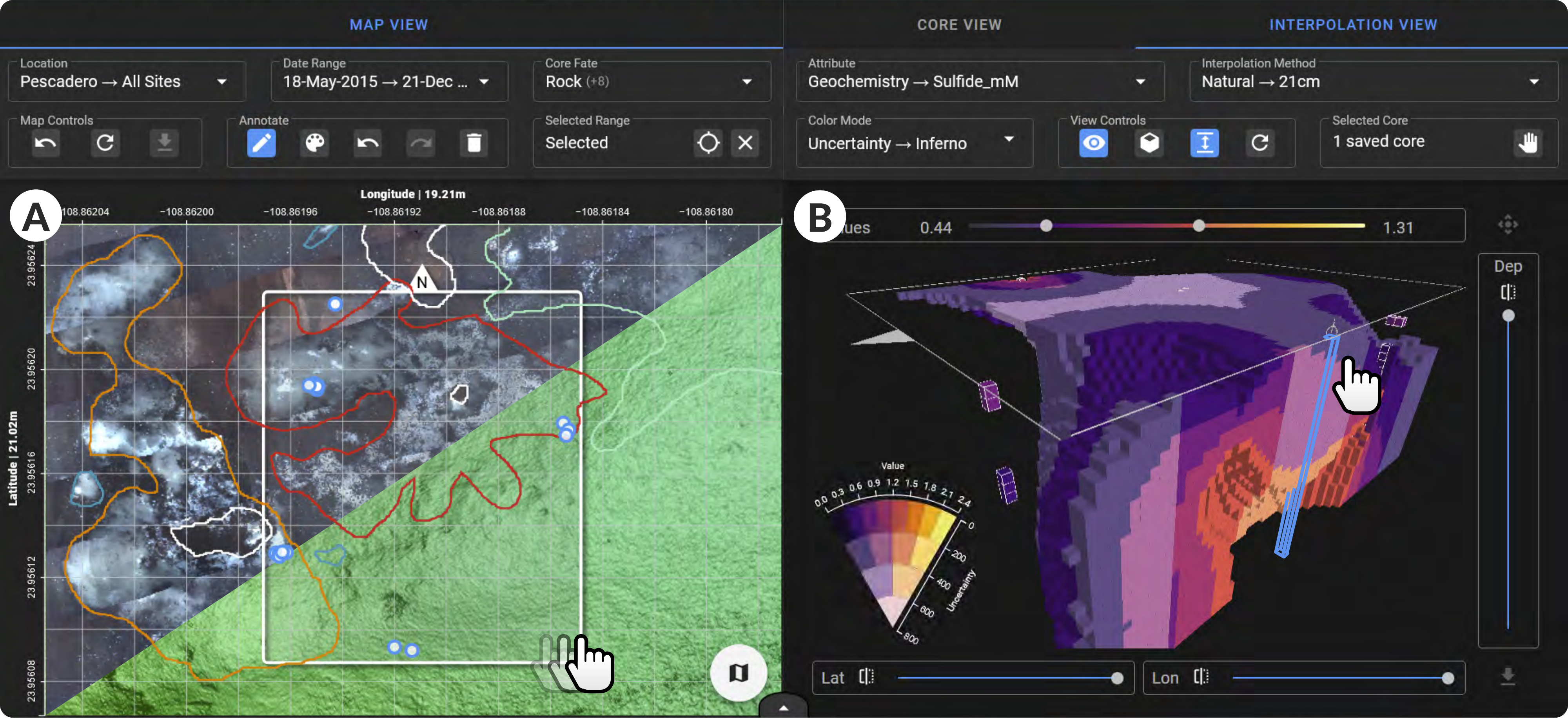}
  \caption{%
    \textit{DeepSee} presents side-by-side views of 2D geological and biological landscape maps \textbf{(A)} as well as 2D visualizations and 3D interpolations of physical, geochemical, and biological parameter gradients in deep sea sediment cores \textbf{(B)}.
  }%
  \Description{%
    The DeepSee web interface.
    From top to bottom, the page consists of a small full-width header with three navigation tabs, a larger full-width section with dropdown menus and buttons, and a main content area with two half-width columns.
    The left column labeled (A) shows the Map View with two map types and user-drawn colored lines as annotations.
    The right column labeled (B) shows the Interpolation View with a colored undersea landscape made up of individual cubes.
  }%
  \label{fig:teaser}
\end{figure*}

Exploration and research on deep-sea benthic ecosystems, ranging from the expansive abyssal plains to the physico-chemically heterogeneous extreme environments of hydrothermal vents and methane seeps, has significantly advanced in the past two decades.
These advances arise from massive amounts of data produced by the development of new technologies for autonomous seafloor mapping and imaging, physico-chemical sensing, biological measurements, and sample archival \cite{Martin:2016:NewTechniquesHydrography}.
Through multidisciplinary teams of scientists providing complementary data sets on the geology, geochemistry, macro- and microbiology of these remote ecosystems, deep ocean researchers can now access 2D topographic maps and color photomosaics of the seafloor, allowing for the relation of point-source seafloor sample collections (e.g. sediment cores, rock, animal, and water samples) with their appropriate environmental context at centimeter to kilometer spatial scales.

However, while 2D maps of spatial locations of samples are valuable, the field currently lacks visualization tools which extend into the 3rd dimension, i.e. within the subseafloor.
This presents several fundamental data visualization challenges around exploring associated physico-chemical and biological data of samples within their spatial context.
The samples are sediment cores -- pointwise geospatial and temporal (i.e., multidimensional) data consisting of hundreds of microbial species and physio-chemical parameters -- that are spread sparsely across several hundred meters of seafloor.
Researchers perform fieldwork to collect these sediment cores from the deep ocean during expeditions, an expensive and time-consuming task that requires extensive planning beforehand and usually results in a very limited number of samples.
Seeing prior sampling history to decide where to sample next can help maximize time spent conducting expensive fieldwork.
Further, it is unclear how scientist's existing research workflows might change by introducing new visualization capabilities, both in planning before fieldwork as well as in tactical decision making in the field as new samples are collected.
Our goals are to develop a system that solves the visualization challenges in exploring spatial trends between sediment cores in the context of their environment, deploy the system to measure the impacts on scientists' workflows conducting deep ocean fieldwork, and reflect on broader design guidance for visualizing multidimensional sample data collected during fieldwork expeditions.

We present a collaborative design study \cite{Sedlmair:2012:DesignStudyMethods} conducted with a team of scientific researchers studying deep ocean ecosystems (Sect.~\ref{sec:methodology}).
We first characterized their research goals, preexisting workflows and data practices, leading us to synthesize design challenges and user tasks (Sect.~\ref{sec:design}) for a system that can help scientists visualize spatial trends between cores in context of the environment, integrate core and map data at multiple size and time scales in a single interface, and increase the scientific return (long-term usability and value of the data) from the limited number of samples available by predicting unseen values in unsampled locations.

From our task abstraction, we developed \textit{DeepSee} (Fig.~\ref{fig:teaser}, Sect.~\ref{sec:system}), an open-source\footnote{\textit{DeepSee} code: \url{https://github.com/orphanlab/DeepSee}}, interactive workspace for scientists to upload sediment core data and map images and see their sampling history displayed across multiple connected views simultaneously.
Interactive maps of the seafloor between centimeter and kilometer resolution (Fig.~\ref{fig:teaser}A) are labeled with information about previous dives as well as collected samples.
Alongside these maps, \textit{DeepSee} displays 2D visualizations that show parameter gradients as a function of depth and interactive 3D visualizations of data interpolations in the space between samples (Fig.~\ref{fig:teaser}B).
The data interpolations can be run in real time, allowing scientists to ``see'' below the seafloor and determine the most likely places to collect high-value samples.
To support decision making, \textit{DeepSee} provides annotation tools on the maps for taking notes, useful for communicating findings and planning future dives.
Finally, \textit{DeepSee} is portable and requires no internet access, empowering scientists to use \textit{DeepSee} on field expeditions in remote environments.

To assess the impact of our approach on solving a real-world problem, we evaluated \textit{DeepSee} with the same expert collaborators involved in the design process (Sect.~\ref{sec:evaluation}).
First, we conducted an initial field deployment of \textit{DeepSee} aboard a research cruise to test its capabilities and illustrate which user tasks were accomplished in the field.
Then, we interviewed each team member and collected feedback on the deployment around several areas of deep ocean fieldwork research that \textit{DeepSee} impacted.
We learned that:

\begin{itemize}
    \item[\textbf{(1)}] Fluid interaction \cite{Elmqvist:2011:FluidInteraction} and aggregation of data visualized at multiple scales can help scientists increase the scientific return of limited samples by enabling new ways of asking questions about the data;
    \item[\textbf{(2)}] Integrating 2D and 3D data in the same interface can catalyze new research workflows by surfacing new potential targets for future sampling and reducing the long-term cost of data preparation; and
    \item[\textbf{(3)}] Modular visualizations can help team members with different roles and diverse research goals to solve specific tasks as they decide where and which samples to collect.
\end{itemize}

From these findings, we synthesized principles for designing future visualization systems in other fieldwork-driven domains:

\begin{itemize}
    \item[\textbf{(1)}] Prioritize data integration as a user task when designing, helping users track data being added on the fly in the field;
    \item[\textbf{(2)}] Visualize physical data in context of the environment, fostering stronger understanding and communication of insights;
    \item[\textbf{(3)}] Combine data types to bridge micro and macro scale analysis, increasing scientific return on limited samples; and
    \item[\textbf{(4)}] Design interactive visualizations to aid mental modeling, helping scientists understand and explore complex processes.
\end{itemize}

Overall, the combination of fluid interaction, tightly integrated data, and modular visualizations provided deep insight into the data that helped researchers build a stronger intuition about the spatial distributions of environmental sample data, ultimately facilitating more informed future sample selection.

In summary, the contributions of this paper are: (1) a characterization of the scientific process and design considerations for visualizing deep ocean ecosystems following a user-centered design process; (2) \textit{DeepSee}, an interactive workspace that visualizes the history of deep sea sediment sampling for selecting future sample site locations; and (3) reflections based on interviews with scientists that deployed \textit{DeepSee} discussing the lessons learned and broader impacts of leveraging data visualizations to support fieldwork-driven scientific research.

\section{Related Work}
\label{sec:related_work}

\noindent\textbf{Scientific Visualization Techniques and Systems.  }
Drawing inspiration from a rich history of scientific visualization challenges in analyzing environmental data \cite{Lipsa:2012:PhysicalSciencesVis, Afzal:2019:SOTAOceanDataVA}, \textit{DeepSee} leverages a combination of multidimensional- and geo-visualization techniques to help scientists contextualize and explore multi-field data.
To bridge the gap between understanding global phenomena and local subsets of interest (i.e. ``details within context'' \cite{Lipsa:2012:PhysicalSciencesVis}), we utilized selection and details-on-demand interactions to help users navigate geo-spatiotemporal visualizations of sampling history at multiple physical and temporal scales \cite{ANDRIENKO:2003:GeospatiotemporalReview, Andrienko:2006:GeospatiotemporalAnalysis, Dykes:2005:ExploringGeoVis}.
Once a local sample subset has been identified, \textit{DeepSee} presents multi-field visualization techniques for comparison \cite{Gleicher:2018:VisComparison} at the data level in both 2D using juxtaposed bar charts and 3D using basic volume rendering techniques \cite{Drebin:1988:VolumeRendering, Song:2006:AtmosphericVA, Hansen:2011:VisualizationHandbook}.
We further leveraged Value-Suppressing Uncertainty Palettes \cite{Correll:2018:VSUP} to represent uncertainty as color for making sense of our simulated data interpolations in 3D.
Our science team collaborators were familiar with GIS (Geographic Information Systems) such as ArcGIS\footnote{\url{https://www.arcgis.com/index.html}} in their work, and compared them to the capabilities in \textit{DeepSee}.
However, while ArcGIS provides state-of-the-art capabilities for lab-based geo-spatiotemporal data analysis, it lacks out-of-the-box support for rapid data integration and visualization of point-based field sample data interpolations in 3D in context of their 2D geolocations.
We aim to bridge this gap in the context of deep ocean fieldwork operations by combining on-the-fly 3D data interpolation capabilities as samples become available with fluid interactions \cite{Elmqvist:2011:FluidInteraction} across side-by-side 2D and 3D views of the data, useful in both planning before fieldwork as well as for tactical decision making in the field.
Our design work seeks to explore ways of interacting with and combining geovisualization and 3D data visualizations in a single interface, towards designing future fieldwork data visualization workflows.

\medskip
\noindent\textbf{Visualization Systems for Analyzing Oceanographic Data.  }
Prior work in developing systems for visually analyzing and predicting ocean data spans multiple diverse use cases, such as forecasting ocean currents for oil exploration \cite{Hollt:2013:VAOceanForecasts} and comparing multiple geospatial and temporal ocean simulation models at once \cite{Kothur:2014:VAOceanModelOutput}.
One of the most widely used tools in the community is Ocean Data View \cite{Schlitzer:2022:OceanDataView}, a general-purpose package for exploring multiparameter profile or sequence oceanographic data.
Other tools present more specific task-based visualizations.
For example, Ovis \cite{Hollt:2014:Ovis} is a visual analytics framework for investigating multivariate sea surface height data collected over time, while SeaVizKit \cite{Wael:2019:SeaVizKit} is a recent toolkit featuring interactive capabilities for visualizing ocean field data at multiple scales in the browser, similar to our browser-based approach.
However, these systems and most others focus on visualizing ocean models, vector-based numerical models of ocean properties such as currents and their circulation.
While we share many of the same challenges including the need for multidimensional views, quantifying uncertainty and supporting both scientific investigation and decision making \cite{Wael:2019:SeaVizKit}, our collaborators' task is to analyze point-based fieldwork samples collected from deep ocean sediments, which introduces new challenges in integrating 2D and 3D data as well as supporting fluid interaction \cite{Elmqvist:2011:FluidInteraction} between these data at multiple spatial and temporal scales.
Further, these systems introduce design principles for use during long-term, lab-based strategic analysis of data, while deep ocean researchers conduct expensive fieldwork that requires support for on-the-fly decision making.
We build on these principles in \textit{DeepSee}'s design by additionally supporting short-term, field-based tactical analysis of data during expeditions.
By developing visualizations that can represent point-based sample data in context of the environment in multiple dimensions, we aim to learn broad principles for designing future visual analytics systems that support fieldwork-driven research.

\medskip
\noindent\textbf{Visualization Design Studies in Fieldwork Contexts.  }
Data visualization often engages with a broad range of disciplines through a design study methodology \cite{Sedlmair:2012:DesignStudyMethods}, a problem-driven research framework used to systematically apply visualization techniques to an application area and evaluate the solution with real users.
Examples of visualization design studies can be found in domains such as energy analysis and modeling \cite{Goodwin:2013:UCDEnergyVis}, cybersecurity \cite{Mckenna:2015:UCDVizSec}, and manufacturing \cite{Eirich:2022:IRVINE}.
Our design study sought to combine data visualization and human-centered design in the context of scientific fieldwork operations, or processes that are time-sensitive, action-oriented, and involve assessing risk and impact across multiple scenarios \cite{Conlen:2018:DesignVAOperations}.
There is a growing body of literature on design principles for building visual analytics systems in the context of science operations that stresses human-centered design and information visualization as both the method and outcome, helping scientists gain insight into their own workflows by seeing the problem through a different lens \cite{Conlen:2018:DesignVAOperations, Hendrie:2022:CollabMethodSciVis}.
For example, MOSAIC Viewer \cite{Bae:2020:VAMultiRobotWorldviews} is a visual analytics system that utilized a human-centered design methodology to help human operators make sense of multiple robots' scheduling decisions and synchronization, while reducing the time and difficulty associated with previous operations.
Specifically in fieldwork-driven ocean science research, \textit{DeepSee} seeks to expand on human-centered approaches to geovisualization design \cite{Lloyd:2011:HCDGeoVis} that put scientists in the driver's seat for interacting directly with the algorithms that enable data exploration and analysis.

\section{Methodology}
\label{sec:methodology}

In this paper, we followed Sedlmair et al.’s methodological framework for design studies \cite{Sedlmair:2012:DesignStudyMethods}.
Each phase of the study resulted in a contribution, namely: (1) a characterization of the problem domain via design challenges and user tasks; (2) a novel and validated visualization design; and (3) reflections on the process and lessons learned for improving visualization and interaction design guidelines.
We connect these phases throughout the paper by illustrating how our synthesized challenges and tasks informed \textit{DeepSee}'s design and how \textit{DeepSee}'s success in addressing the challenges and tasks is reflected in our evaluation.
To the best of our knowledge, the problem we describe in Sect.~\ref{sec:design} is unique to our collaborators and has not been addressed by other researchers so far.
Thus, we posit that a design study based on active participation from collaborators in both the design and evaluation phases is necessary to successfully support domain experts in their highly specialized tasks.

We began by establishing a close collaboration between visualization researchers and domain experts, to properly assess the impact of our approach on solving a real-world problem. 
In line with prior design studies \cite{Goodwin:2013:UCDEnergyVis, Mckenna:2015:UCDVizSec, Eirich:2022:IRVINE}, our team comprised all authors and included visualization experts in human-centered computing, interaction design, scientific data visualization and art, as well as scientists and researchers with expertise in environmental microbiology, geochemistry and geology.
We developed \textit{DeepSee} through several user-centered design methodologies including contextual inquiry, mixed-fidelity prototyping, user testing, and design iteration.
Over an intensive ten-week period, we worked closely together to develop a shared understanding of the importance and aims of their scientific research (Sect.~\ref{sec:scientific_value}), characterize data and workflows (Sect.~\ref{sec:workflow_and_data}), and identify data visualization challenges and user tasks (Sect.~\ref{sec:design_challenges}) that would guide our system design.

Our process started with contextual inquiry, combining observation and interviews with our domain collaborators to deeply understand their existing tools and workflows, discuss unmet needs and pain points, and identify key technical constraints.
After two weeks, we began mixed-fidelity prototyping, starting with paper prototypes, wireframes and small interactive prototypes in code, before moving on to progressively higher-fidelity prototypes.
These prototypes were iteratively discussed with the science team over the next seven weeks through semi-structured user testing.
Using these prototypes as stimuli, we deepened our understanding of the problem space through on-going contextual inquiry.
During this time, systems diagrams and conceptual workflows were also created to help us validate our understanding of the problem space in collaboration with the science team. 
As a result of this approach, we progressively uncovered insights that guided each subsequent design iteration, leading us to generate a working prototype described in Sect.~\ref{sec:system} that addressed our collaborators' tasks.

\textit{DeepSee} was then deployed \textit{in the wild} \cite{Sedlmair:2012:DesignStudyMethods} to assess whether our approach solves a real problem -- bridging gaps in supporting visual analysis of samples during field research.
Deploying \textit{DeepSee} with the same experts that we designed with allowed us to assess whether domain experts are indeed helped by the new solution, as well as reflect on and refine principles for visualization and interaction design guidelines \cite{Sedlmair:2012:DesignStudyMethods}.
Specifically, we investigated the impact of introducing data visualizations into both planning before fieldwork as well as tactical decision making in the field.
We learned that visualizing data at multiple scales, integrating 2D and 3D data in the same interface, and modularizing visualizations positively impacted scientists planning workflows the most, and that additional support for integrating data on the fly could better enable in-situ decision making.
Methodological details of our evaluation and how we synthesized these lessons learned are provided in Sect.~\ref{sec:evaluation}.

\section{Studying Deep Ocean Ecosystems}
\label{sec:design}

As the first step in the design study methodology, we characterized the needs of deep ocean researchers when visualizing multidimensional environmental data collected during field expeditions.
What makes this challenging is addressing the size and complexity of data collected in a single interface, where researchers can correlate processes at several size scales, explore spatial trends in context of the environment, and test hypotheses where samples are sparse and limited.
\textit{DeepSee} addresses these challenges specifically in studying deep ocean ecosystems by integrating multiple coordinated views that both present data at multiple size scales simultaneously and visualize geochemical gradients and variation in microbial taxa in the context of the environment in both 2D and 3D.
This can help researchers decide on new sampling locations that maximize scientific return.

\subsection{Deep Ocean Research Goals}
\label{sec:scientific_value}
Broadly, our collaborators seek to advance an understanding of deep ocean microbial ecosystems.
The deep ocean is Earth's largest biome and is home to a variety of habitats that host rich, complex chemosynthetic ecosystems including hydrothermal vents, cold seeps, gas hydrates, whale falls, and carbonate platforms \cite{Corinaldesi:2015:PerspectivesDeepSeaEcology}.
Microorganisms represent primary drivers of these extreme ecosystems, supporting diverse animal communities \cite{Jorgensen:2007:FeastAndFamine, Danovaro:2014:ParadigmsDeepSeaEcology}.
Importantly, these deep ocean microbial ecosystems have been observed to facilitate key environmental processes, including the transformation of greenhouse gases such as methane and carbon dioxide, movement of energy between trophic levels, and general cycling of elements, such as sulfur and metals, making these ecosystems critical drivers on the global ecological scale \cite{Nealson:1997:SedimentBacteria, Mason:2009:ProkaryoticDiversity}.

Clearly defining the nature, role, and impact of microorganisms in deep ocean ecosystems is critical for connecting deep ocean processes to biogeochemical cycling and diversity globally.
There is increased importance of temporal and spatial monitoring with exploitation of deep ocean resources (e.g. deep sea mining) or in response to environmental change.
Characterizing these extreme environments and the microbial adaptations required to persist and carry out key ecological processes is also advantageous for applications in the industrial world, where yet-to-be discovered molecular machinery and compounds may be applied to a variety of unsolved problems in medicine, agriculture, and environmental science \cite{Ohta:2006:Carrageenase, Wu:2006:CharacterizationXylanase, Cavicchioli:2011:UsesOfPsychrophiles}. 
Similarly, advances in planetary exploration have also revealed similar environments to those found in Earth's deep oceans; further exploration of Earth's deep ocean microbial ecosystems can therefore further shape our understanding of how life can adapt to the similarly extreme conditions that may exist beyond Earth \cite{Hand:2009:AstrobiologyLifeOnEuropa, Martin:2018:LifeOnETOceanWorlds, MacKenzie:2022:EvidenceLifeAtEnceladus}.

\subsection{Workflow and Data}
\label{sec:workflow_and_data}

\begin{figure}[!t]
  \centering
  \includegraphics[width=\linewidth]{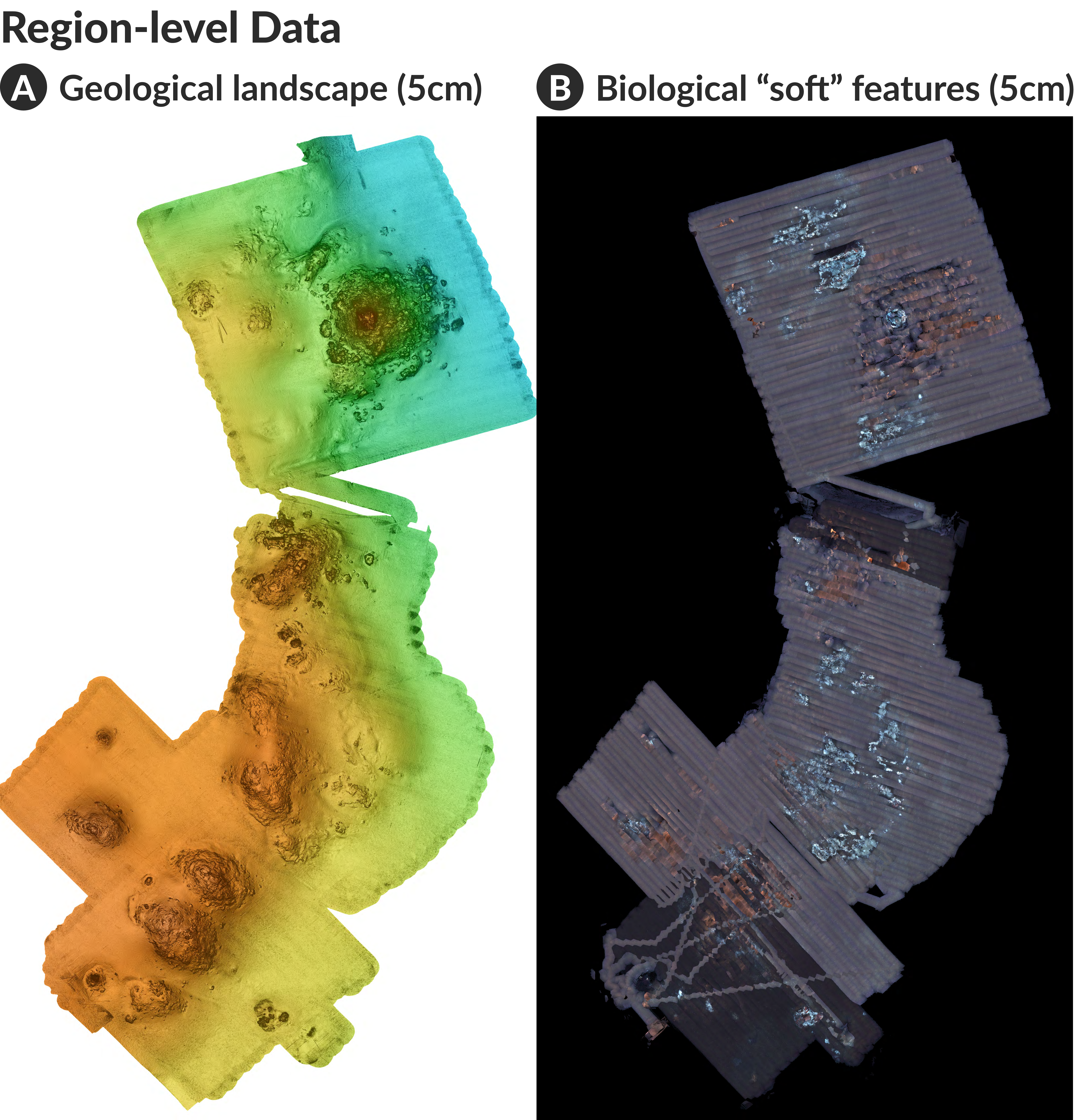}
  \caption{%
    Region-level data represented as maps, provided by \cite{Paduan:2022:VentMaps}.
    Maps reveal two important features: \textbf{(A)} geological landscapes, such as locations of hydrothermal vent mounds and chimneys, visualized by topography from AUV and low-altitude survey system (LASS) multibeam surveys; and \textbf{(B)} surface biological and ephemeral “soft” features, such as white patches of microbial mats growing in hydrothermal fluids, visualized by photomosaics, acoustic backscatter and LIDAR bathymetry.
    \textit{DeepSee} solves a critical challenge of combining region-, core-, and sample-level data in a single interface using expressive and interactive visualizations (Sect.~\ref{sec:system}).
  }%
  \Description{%
    Two photos labeled (A) and (B) show maps of the ocean floor.
    Image (A) shows a geological landscape colored by elevation.
    Image (B) shows a photo mosaic of the same landscape.
  }%
  \label{fig:region_data}
\end{figure}

\begin{table*}
  \caption{%
    A Subset of Core-Level and Sample-Level Data Visualized by \textit{DeepSee}\protect\footnotemark
  }%
  \label{tab:core_sample_data}
  \begin{tabular}{r *{10}{l}}
    \toprule
      & \multicolumn{6}{l}{Core \textit{(10+ identification and measurement params)}} & \multicolumn{4}{l}{Sample \textit{(500+ phys/geo/bio params)}} \\
    \midrule
      Core ID & Location & Date & Core Fate & Latitude & Longitude & Horizon & Sulfate & Sulfide & Taxa 1 & Taxa 2 \\
    \midrule
      NA091\_020 & Auka - Matterhorn & 11-01-17 & Geochem & 23.954198 & -108.862394 & 2-3 cm & 22.98 & 5.14 & 0.1358 & 0 \\
      NA091\_020 & Auka - Matterhorn & 11-01-17 & Geochem & 23.954198 & -108.862394 & 3-4 cm & 17.97 & 4.6 & 50.497 & 0 \\
      ... & ... & & & & & & ... & & &  \\
      S0193\_PC5 & Auka - Diane's vent & 11-14-18 & Geochem & 23.954822 & -108.863020 & 1-2 cm & 8.85 & 3.78 & 0.4464 & 37.1574 \\
      ... & ... & & & & & & ... & & &  \\
    \bottomrule
  \end{tabular}
\end{table*}

To quantify the distribution of microbial activity in deep sea sediments, microbial ecologists, geochemists and geologists first collect biological and geological samples (often sediment, rock or pore water) from the deep ocean.
These samples are then used to examine the geochemical, sedimentological, and biological characteristics of the environment.
The sampled cores and associated parameters are both sparse and multidimensional (i.e., cores with a fixed location in 3D, sampled at time points ranging from days to decades).

Collecting samples begins with an oceanographic research \textbf{expedition}, where scientists use various human- or robot-operated tools to map the study area and execute sampling dives.
For the examples presented in this paper, AUVs (Autonomous Underwater Vehicles) equipped with multibeam sonar collected 1 m lateral resolution bathymetry and acoustic backscatter covering an $\approx30$ km$^2$ area including multiple hydrothermal vent fields.
With this context mapping completed, a Low Altitude Survey System (LASS) fielded on an ROV (Remotely Operated Vehicle) then mapped one vent field at much higher resolution, yielding 5 cm resolution bathymetry from a multibeam sonar (Fig.~\ref{fig:region_data}A), 1 cm resolution bathymetry from LIDAR, and 5 cm photomosaics from a color stereo camera rig (Fig.~\ref{fig:region_data}B) \cite{Paduan:2022:VentMaps}.
With these maps, scientists then planned the exploration and execution of sampling dives with ROVs or HOVs (Human Occupied Vehicles) to collect sediment samples from the ocean floor and bring them back to the ship for downstream processing.
The samples \textit{DeepSee} currently visualizes are sediment \textbf{push cores (PC)}, hollow polycarbonate tubes 7 cm diameter and 30 cm tall, usually sectioned shipboard into 1 cm or 3 cm tall \textbf{horizons}.
There are two types of data attributes: (1) core-level (Table~\ref{tab:core_sample_data}, left) and sample-level (Table~\ref{tab:core_sample_data}, right).
For core-level analysis, there can be more than a dozen separate measurements conducted for a single horizon with a core; considering a core can have between 7-12 horizons, there are over 84 identification data points to interpret per core.
For sample-level analysis, each horizon is processed for a wide variety of geochemical \textbf{parameters}, including sulfide, sulfate, iron, and other \textbf{chemicals} associated with metabolism or the state of the environment.
Each horizon is also preserved and sequenced for a quantitative measure of how many microbial community members (\textbf{taxa}, e.g., ANME archaea or sulfate-reducing Desulfobacterota) live there.
This typically results in over $500$ discrete taxonomic, geochemical, or physical parameters per core, used to identify spatial trends in distributions (Sect.~\ref{sec:design_challenges}).
Finally, cores are often collected in pairs or clusters for different experiments and assigned a \textbf{core fate} descriptor, e.g., ``Geochem'' or ``Live'', which differentiates what type of metadata and parameters are associated with the core.
In \textit{DeepSee}, these data were stratified into three hierarchical levels and visualized in different ways:

\medskip
\noindent\textbf{Region-level data.  }
Region-level data (Fig.~\ref{fig:region_data}) are predominantly map-format, including low-altitude survey system (LASS) multi beam amplitude and bathymetry (5 cm) \cite{Martin:2016:NewTechniquesHydrography}, AUV bathymetry (1 m lateral resolution), photo mosaics (5 cm resolution), and LIDAR bathymetry (2 cm).
Together, these data types reveal the geological landscape (bathymetry) and surface biological and ephemeral “soft” features (photo mosaics and high-resolution LIDAR).

\medskip
\noindent\textbf{Core-level data.  }
Core-level data (Table~\ref{tab:core_sample_data}, left) are unique x,y points within a methane seep where subsurface sediments were sampled.
Cores comprise a minimum of 11 parameters stored in a CSV file, including core IDs (e.g., dive number, core number, GPS location), measurements (e.g., temperature, sulfate, sulfide), and associated data (e.g., core type, habitat description).

\medskip
\noindent\textbf{Sample-level data.  }
Sample-level data (Table~\ref{tab:core_sample_data}, right) is typically parameters subdivided within cores into horizons of 2-3 cm each.
\textit{DeepSee} treats each horizon (row) as a sample belonging to a core within a seep.
Recent cruises have returned cores with up to 257 depth horizons described in CSV format by 21 physicochemical parameters (e.g. pH), 12 geochemical measurements (e.g. methane concentration), and over 500 biological parameters (microbial identity and abundances).

\footnotetext{The example data is provided by Speth et al. \cite{Speth:2022:MicrobialAukaVents}.}

\subsection{Design Challenges and User Tasks}
\label{sec:design_challenges}

\begin{table}
  \caption{%
    Design Challenges (C) and User Tasks (T)
  }%
  \label{tab:challenges_tasks}
  \begin{tabular}{rl}
    \toprule
      \textbf{C1} & Observing spatial trends in sparse, tabular data \\
    \midrule
      \textbf{T1} & Correlate taxonomic profiles against geochemistry \\
      \textbf{T2} & Estimate spheres of influence around a seafloor feature \\
      \textbf{T3} & Annotate relationships between environmental features \\
      \textbf{T4} & Compare horizons across multiple cores simultaneously \\
    \midrule
      \textbf{C2} & Integrating data at multiple scales of size and time \\
    \midrule
      \textbf{T5} & Contextualize scales between tables and maps \\
      \textbf{T6} & Select cores across expeditions and date ranges \\
      \textbf{T7} & View data at multiple spatial resolutions \\
    \midrule
      \textbf{C3} & Maximizing the scientific value of limited sampling \\
    \midrule
      \textbf{T8} & Choose sample sites likely to address science team goals \\
      \textbf{T9} & Compare past geochemical/microbial data with new \\
      \textbf{T10} & Correlate visual indicators with microorganism data \\
      \textbf{T11} & Estimate unseen parameters in unsampled locations \\
    \bottomrule
  \end{tabular}
\end{table}

By understanding the research goals of studying deep ocean ecosystems (Sect.~\ref{sec:scientific_value}) as well as scientists' workflows and data (Sect.~\ref{sec:workflow_and_data}), we synthesized several key design challenges \textbf{(C)}.
Throughout, we highlight specific user tasks \textbf{(T)} that we aim to support.
The challenges and tasks are summarized in Table~\ref{tab:challenges_tasks} and integrated into our descriptions of the system (Sect.~\ref{sec:system}), usage scenarios (Sect.~\ref{sec:usage}) and evaluation (Sect.~\ref{sec:evaluation}) throughout the rest of the paper.

\medskip
\noindent\textbf{C1  }
\textbf{Observing spatial trends in sparse, tabular data.}
Characterizing the general geochemical and taxonomic data associated within a core is difficult.
Using multiple parameters measured for each horizon, scientists seek to understand sediment habitats of vent and seep microorganisms by \textit{correlating taxonomic profiles against various geochemical parameters} \textbf{(T1)} within a core.
For example, the interplay between sulfate and sulfide concentrations (geochemical parameters) could be compared to microbial members that oxidize sulfate to sulfide (biological parameters).
With these characterizations, scientists can then \textit{estimate spheres of influence  around a seafloor feature} \textbf{(T2)} to answer specific questions such as ``Does community distribution and geochemistry change with increasing distance from the target area of sampled push cores?''
Table~\ref{tab:core_sample_data} provides an example of the variety of data associated with a single sediment core, including multiple types of geochemical measurements, relative abundance measurements of microbial taxa, positional coordinates, depth into the seafloor, sediment temperature, etc.
Our interface should enable users to explore/highlight trends in their data by \textit{annotating relationships between environmental features} \textbf{(T3)} as well as \textit{showing several sediment horizons down the profile of multiple cores simultaneously} \textbf{(T4)}, rather than in solidarity.

\medskip
\noindent\textbf{C2  }
\textbf{Integrating data at multiple scales of size and time.}
The context of the environment around a core also includes additional metadata that is difficult to integrate between time scales, size resolutions, and file formats.
It is critical to \textit{understand relationships between centimeter-scale core data in tables and kilometer-scale seafloor features in maps and images, as well as across scientific research expeditions over long- and short-term scales} \textbf{(T5)}.
For example, rows in Table~\ref{tab:core_sample_data} can span multiple expeditions over several years, sample parameters at each horizon can take several months to measure in the lab, while some environmental metadata can be obtained in real time during a dive.
These tables only capture core-level and sample-level data; region-level data (Fig.~\ref{fig:region_data}) from image files contains scale, location, and depth information that can contextualize cores in a 3D environment.
To integrate these data, our system should allow users to \textit{select and view cores across expeditions and date ranges} \textbf{(T6)}, as well as \textit{show tabular core data in relation to map images with associated data at multiple spatial resolutions} \textbf{(T7)}.

\medskip
\noindent\textbf{C3  }
\textbf{Maximizing the scientific value of limited sampling.}
Biological sampling faces a difficult economy -- the deep ocean is remote, requiring expensive ship and ROV time, and typically only a small number of cores ($\approx20$) can be processed per dive.
Even for teams of experienced scientists, \textit{choosing sampling locations most likely to address key goals of the science team} \textbf{(T8)} is a challenging problem.
For example, during a dive at methane seeps, scientists are limited to what they can see on the surface of the seafloor to make decisions about where to sample.
In some locations, the microbes can form microbial ``mats'' at the seabed, creating visual surface indicators of underlying methane seepage.
However, these indicators are like ``the tip of the iceberg'' with limited ability to predict the diversity of microorganisms within the sediment from the surface.
Sampling is also limited by the ROV's or HOV’s physical capacity to collect and return cores each dive, making it a requirement for scientists to know their data and scientific objectives well.
Thus, selecting the best places to sample can be supported by \textit{analyzing geochemical or microbial data from past samples} \textbf{(T9)} and \textit{observing how visual indicators may correlate with subsurface microorganisms} \textbf{(T10)}.
It is essential to provide multiple capabilities within our system for \textit{estimating unseen parameters from the data tables in unsampled locations} \textbf{(T11)} through spatial association with visual indicators captured by bathymetric and photomosaic map data.

\section{The \textit{DeepSee} System}
\label{sec:system}

From our design challenges and user tasks, we developed \textit{DeepSee}, an interactive workspace for studying deep ocean ecosystems.
Our system tightly integrates three coordinated views.
In the \textit{Map View} (Sect.~\ref{sec:map_view}), users explore maps at near-centimeter resolution of the ocean floor overlaid with locations of and information about previous dives and sampled cores.
After selecting cores of interest, users can compare geochemical, physical, and biological attributes across horizons between cores in the \textit{Core View} (Sect.~\ref{sec:core_view}) or analyze interpolated attribute values in 3D between cores in the \textit{Interpolation View} (Sect.~\ref{sec:interp_view}).

\subsection{Map View}
\label{sec:map_view}

\begin{figure}[!t]
  \centering
  \includegraphics[width=\linewidth]{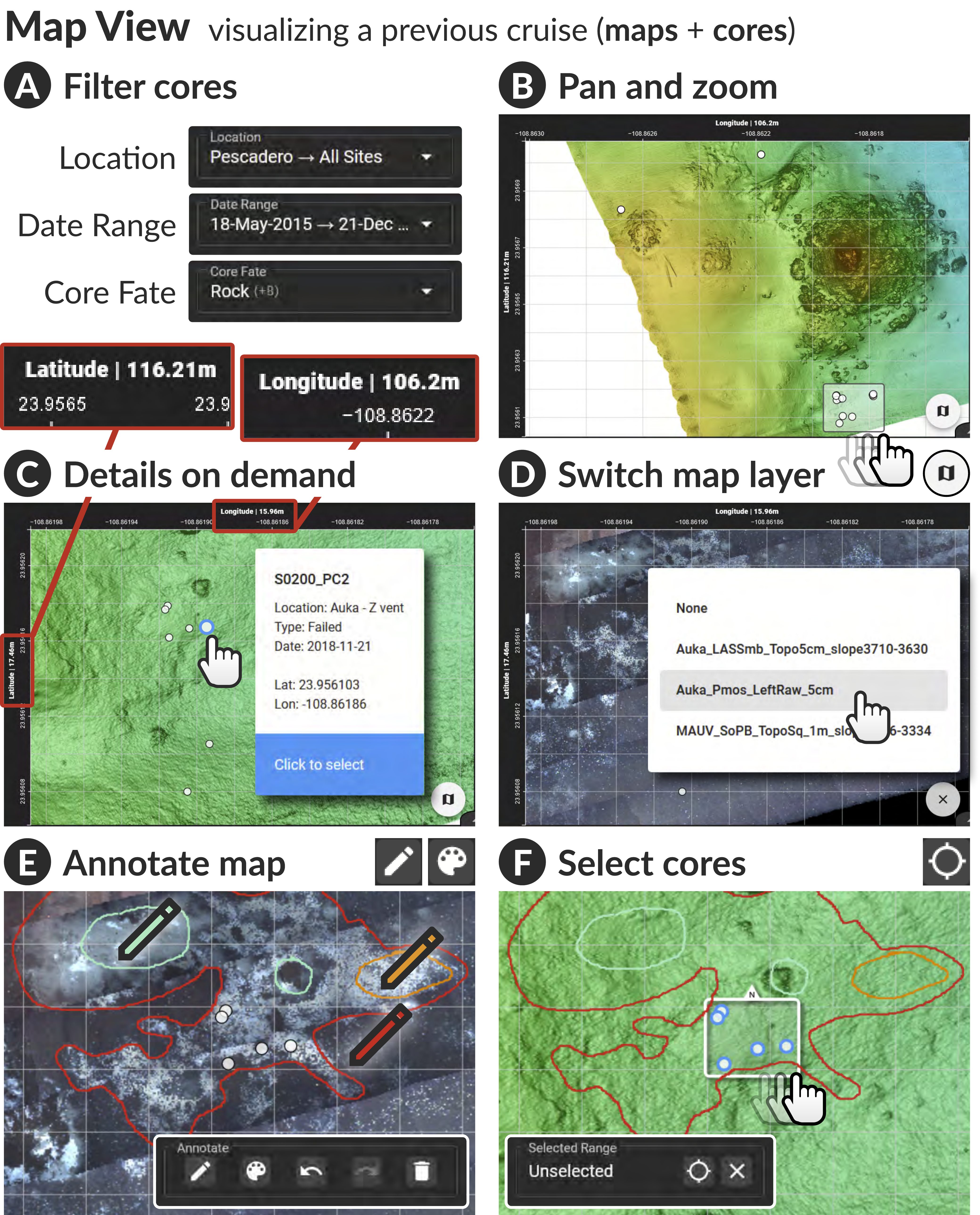}
  \caption{%
    The \textit{Map View} plots cores by latitude/longitude on a map layer to show the spatial and geographic history of sampling.
    Users can drill down to cores of interest \textbf{(A)}, explore the map \textbf{(B)}, see details on demand \textbf{(C)}, switch maps on the fly \textbf{(D)}, draw annotations \textbf{(E)}, and select cores \textbf{(F)} to view in the \textit{Core View} and \textit{Interpolation View}.
  }%
  \Description{%
    Six images labeled (A), (B), (C), (D), (E) and (F) each show the Map View with the same data in different configurations.
    Image (A) shows the dropdown menus for filtering data.
    Image (B) shows a pan and zoom interaction by clicking and dragging a cursor.
    Image (C) shows the results of (B) and shows a tooltip with a cursor hovering on a white dot.
    Image (D) shows a different map and a dropdown menu with a cursor selecting a new map name.
    Image (E) shows the results of (D) and three colored lines drawn by the user.
    Image (F) shows a cursor clicking and dragging over a region of white dots to select them.
  }%
  \label{fig:map_view}
\end{figure}

The \textit{Map View} (Fig.~\ref{fig:map_view}) plots previously sampled cores as a scatter plot using latitude/longitude on top of maps representing geological landscapes (e.g., bathymetry) and surface biological and ephemeral ``soft'' features (e.g., photomosaics and high-resolution lidar).
By visualizing both region-level (map) and core-level (measurement) data simultaneously, users gain more context about spatial trends in the sampling environment by familiarizing themselves with data from previous cruises to help make new inferences about potential future sampling locations.

Users start with filtering cores (Fig.~\ref{fig:map_view}A) by named location, date range, and core fate \textbf{(T6)}.
For example, a scientist not familiar with a previous field site could explore names and date ranges to understand when cores were sampled, or a scientist returning to the site could be interested in a specific geological feature or set of cores previously sampled for a follow-up investigation \textbf{(T9)}.
Then, users navigate the plot by clicking and dragging to pan and zoom (Fig.~\ref{fig:map_view}B) \textbf{(T7)}.
Cores remain the same size at all levels of zoom to avoid screen clutter and help users track them as the view changes.
We also provide several details on demand (Fig.~\ref{fig:map_view}C).
We label the plot with latitude and longitude grid lines and compute the width and height of the current viewport in metric units.
This gives users a baseline for the size of the geological and biological features they see on the map as well as the distance between sampled cores at all times \textbf{(T5)}.
On hover, users can see the named location, fate, date, and exact latitude and longitude of any core.
Users can also switch the map on the fly (Fig.~\ref{fig:map_view}D) and the boundaries of the viewport will persist, helping users maintain context as the view changes \textbf{(T5)}.

Once a site of interest has been located, we built an in-situ drawing tool on the map with six colors and undo/redo capabilities (Fig.~\ref{fig:map_view}E) to support evidence marshaling through annotation \textbf{(T3)}.
For example, a user could draw an outline on a bathymetric map around a surface expression of biological activity, write a small note to remember/share their intent, then switch to a seafloor mosaic image to see the same drawn feature projected on the sediment composition \textbf{(T2)}.
The user naturally concludes their initial exploration by selecting cores (Fig.~\ref{fig:map_view}F) to visualize in the \textit{Core View} or \textit{Interpolation View} \textbf{(T8)}.
After clicking and dragging over any number of points, we draw the rectangular convex hull that encases the selected points as a white border with a triangle at the top that always points North \textbf{(T5)}.
This border is drawn again in the \textit{Interpolation View} to help orient the user at all times when comparing with the \textit{Map View}.

\subsection{Core View}
\label{sec:core_view}

\begin{figure}[!t]
  \centering
  \includegraphics[width=\linewidth]{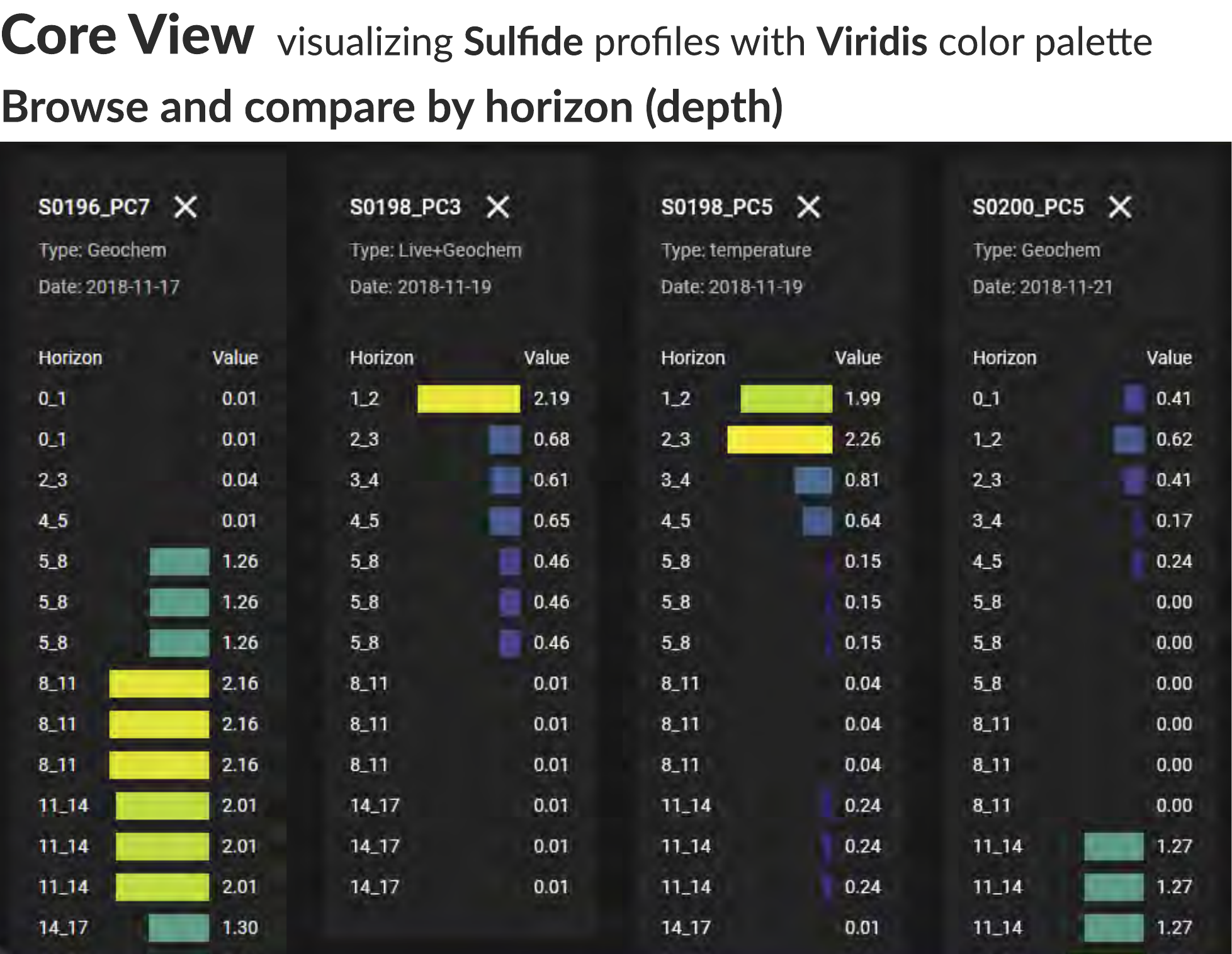}
  \caption{%
    The \textit{Core View} arranges sets of horizontal bar charts to compare parameter values between cores across horizons.
  }%
  \Description{%
    One image shows the Core View.
    Four horizontal bar charts are aligned vertically and arranged horizontally side-by-side.
    They each show measured sulfide values down the profile of a different core sample.
  }%
  \label{fig:core_view}
\end{figure}

The \textit{Core View} (Fig.~\ref{fig:core_view}) arranges columns of horizontal bar charts that visualize a single core-level (measurement) or sample-level (parameter) attribute as individual bars at each horizon of the selected cores in the \textit{Map View}.
This plot enables rapid comparison of attributes in one dimension (depth) to quickly gain insights into biological and geochemical parameter distributions, helping users investigate subsurface gradients or trends among selected cores from a single expedition or multiple expeditions simultaneously.

To add a chart of a single attribute for each selected core in the \textit{Map View}, users search for and select attributes using a drop-down menu.
The bar charts show core IDs, fates, dates, and are aligned vertically by depth horizon to facilitate comparison across bar charts along the same horizontal baseline \textbf{(T4)}.
For example, with a small number of cores selected around a vent in the \textit{Map View}, a user could use the \textit{Core View} to ask, ``Which cores have high sulfide values? Do all cores with high sulfide also have high relative taxa `X' abundance? If not, what other geochemical parameters are different that might explain this difference?'' \textbf{(T1, T2)} leading to predictive analysis such as ``As we move further away from this vent, do we see these values increasing or decreasing? Are all orange mats associated with relative taxa `X' abundance?'' \textbf{(T10)}
We chose horizontal bars to visually mirror the measurement of parameters by depth in the sediment.
Where horizons are different sizes, the smallest step size (typically 1 cm) is used.
If a core does not have sample data at that resolution (e.g., parameters measured every 3 cm), we duplicate the bar by the number of smallest steps at that scale and repeat the horizon label to make this clear for the user.
Users have a choice of six colorblind-friendly palettes (Viridis, Cividis, Greyscale, Inferno, Plasma, Magma).
Finally, users can directly deselect cores from this view, to narrow the space of cores for subsequent analysis in the \textit{Interpolation View} \textbf{(T8)}.

\subsection{Interpolation View}
\label{sec:interp_view}

\begin{figure*}[!t]
  \centering
  \includegraphics[width=\linewidth]{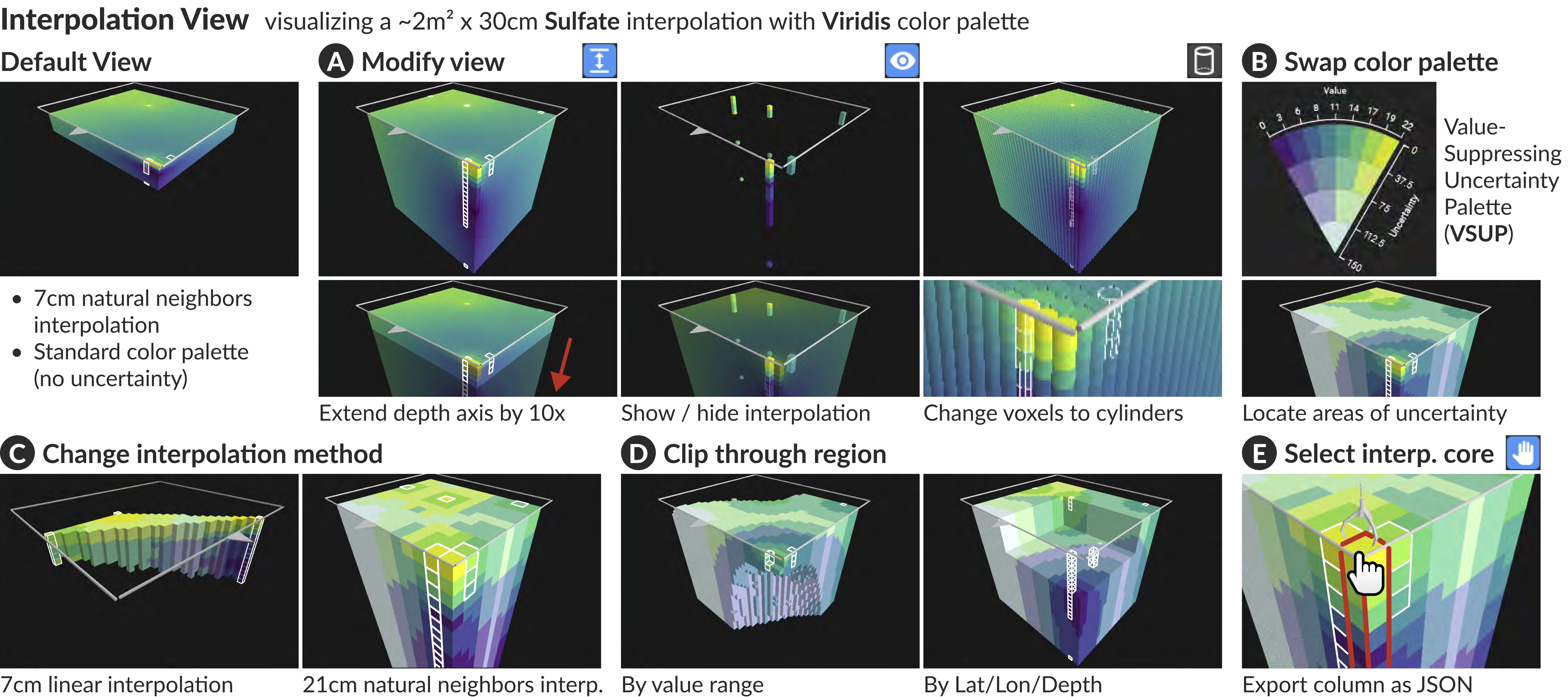}
  \caption{%
    The \textit{Interpolation View} visualizes parameter values interpolated in three dimensions (latitude/longitude/depth) between cores.
    We provide several reconfiguration interactions \textbf{(A)} to help users make sense of an unfamiliar data representation.
    Users can update the view on the fly in several ways: swapping between standard and Value-Suppressing Color Palettes (VSUPs) \cite{Correll:2018:VSUP} \textbf{(B)}; changing the interpolation method and/or grid size \textbf{(C)}; and clipping through the interpolation \textbf{(D)}.
    Finally, users can select an interpolated core to export as JSON \textbf{(E)}.
  }%
  \Description{%
    Fourteen images each show the Interpolation View with the same data in different configurations.
    One image shows a default configuration of an isometric rectangular prism made up of smaller colored rectangular prisms representing interpolated Sulfate values at every latitude, longitude, and depth coordinate.
    Six images labeled (A) show three spatially modifications to the prism with image pairs.
    The first pair shows elongating one dimension of the rectangular prism.
    The second pair shows hiding non-interpolated smaller rectangular prisms.
    The third pair shows each smaller rectangular prism changed into a cylinder.
    Two images labeled (B) show a value-suppressing uncertainty palette legend and the palette applied to the color of the smaller rectangular prisms.
    Two images labeled (C) show two different interpolation method results that change the shape of the rectangular prism.
    Two images labeled (D) show smaller rectangular prisms hidden when the user clips them from the view.
    One image labeled (E) shows a cursor selecting on a column of smaller rectangular prisms representing a single core.
  }%
  \label{fig:interp_view}
\end{figure*}

\aptLtoX[graphic=no, type=html]{
}{
    \begin{figure*}[!t]
      \centering
      \includegraphics[width=\linewidth]{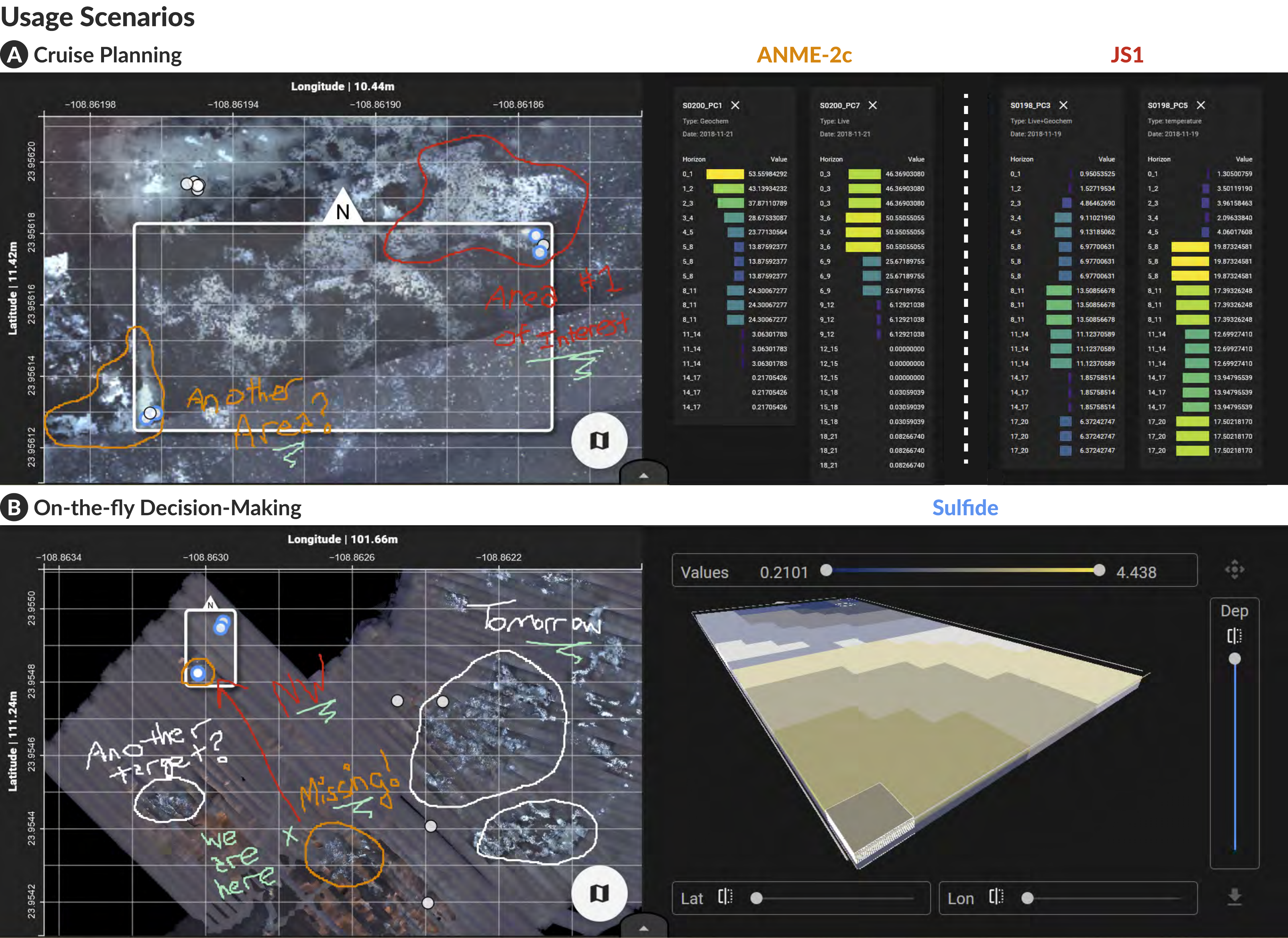}
      \caption{%
        Two usage scenarios showing how \textit{DeepSee} helps users make the most of limited sample data by supporting real-time annotation and interpolation in the tool, with data from Speth et al. \cite{Speth:2022:MicrobialAukaVents}.
        In \textbf{(A)}, our user maximized information between map and tabular data to determine where high-value cores containing both ANME-2c and JS1 are most likely to occur on future dives.
        In \textbf{(B)}, during an on-the-fly decision-making scenario where the seafloor changed over time, the \textit{Interpolation View} created a data-driven opportunity to correct course by interpolating unseen sulfide values and helping users find new targets.
      }%
      \Description{%
        Two images labeled (A) and (B) each showing the half-width double column main content area of DeepSee.
        Image (A) shows the Map View on the left and the Core View on the right.
        The Map View shows colored line annotations and the Core View shows ANME-2c and JS1 values for two core samples.
        Image (B) shows the Map View on the left and the Interpolation View on the right.
        The Map View shows colored line annotations and the Interpolation View shows a colored gradient of sulfide representing high values in the southwest corner transitioning to low values in the northeast corner.
      }%
      \label{fig:usage_scenarios}
    \end{figure*}
}

The Interpolation View (Fig.~\ref{fig:interp_view}) visualizes sample-level (parameter) data by interpolating values between cores in three dimensions (latitude, longitude, and depth) and rendering the resulting data as a three-dimensional object colored by the parameter value at each point in space.
Our visualization technique can extend to any 3D interpolation method that provides four attributes ($x$, $y$, $z$, and a value) for each point observation.
The interpolations can be run in real time, allowing the scientists to rapidly investigate different predictions about the environment below the seafloor where future sampling might yield the greatest return on investment.

\textit{DeepSee} renders 3D interpolation data as space-filling volumetric pixels (voxels) in a 3D space.
For example, at an $x$ cm grid size, we represent an interpolated value at a given latitude/longitude/depth as a rectangular prism with width/length of $x$ cm and a height of 1 cm, matching the same depth horizon measurements in real cores.
We currently support fast 3D linear barycentric interpolation and a 3D discretized approximation of Sibson's natural-neighbors interpolation \cite{Park:2006:NaturalNeighbors} \textbf{(T11)}.
While the research scientists expressed concerns that linear and natural-neighbors gradients are unlikely to reflect spatially heterogeneous in-situ environmental processes, they were optimistic about the capabilities of visualizing these methods as a proof-of-concept.
Linear interpolations offer a simple explanatory model with little complexity and no ability to interpolate outside of sampled regions.
Natural-neighbors is a smoother approximation of nearest-neighbors interpolation with a space-filling property; i.e., outside the space between observations, values will decay at the rate of the gradient at the boundary.
Additionally, the grid sizes offered were specifically chosen to mirror the real life size of cores (7 cm diameter) \textbf{(T7)}.
Because interpolations are computationally expensive, especially at 7 cm resolution across tens to hundreds of meters, we provide larger grid sizes in 7 cm increments.

We used the built-in camera system with OrbitControls from Three.js, the JavaScript library we used to build the Interpolation View (Sect.~\ref{sec:implementation}).
Using a mouse, users can left click and drag to freely rotate the voxels with a stationary camera angle, right click and drag to pan the camera in the latitude/longitude plane, and scroll the mouse wheel to zoom in and out.
We built additional features that allow users to reconfigure the view in several ways \textbf{(T7)} using buttons (Fig.~\ref{fig:interp_view}A): (1) extending the height of each voxel from 1 cm to 10 cm, making patterns in depth much easier to see; (2) showing / hiding all interpolated values and only showing actual sample data; and (3) changing voxels from rectangular prisms to cylinders at a fixed 7 cm diameter, mirroring the shape of real life cores.
These visual aids helped orient new users quickly to this projected visualization method.

The research scientists also expressed an interest in seeing where interpolations were less ``certain''.
In response, we implemented variations on the six color palettes from the \textit{Core View} based on Value-Suppressing Uncertainty Palettes (VSUPs) \cite{Correll:2018:VSUP} (Fig.~\ref{fig:interp_view}B).
Where color palettes apply a color to a given voxel based on a single parameter value, VSUPs take in an additional uncertainty parameter and suppress the color as a function of the amount of uncertainty in the value \textbf{(T11)}.
In \textit{DeepSee}, we normalized the linear distances from each voxel to the nearest non-interpolated voxel as uncertainty; thus, interpolated values farther from a known sample value will appear suppressed, signaling a lack of confidence in the interpolation at that location.
Similar to the interpolation methods, our choice of uncertainty measure (distance to nearest sample) worked as a proof of concept for how to integrate more scientifically accurate measures in the future.
The color palette can be changed on the fly while preserving the camera angle and visual embellishments.

We provide several on-the-fly capabilities to help users explore interpolations while maintaining spatial context.
Users can change the interpolation method and grid size (Fig.~\ref{fig:interp_view}C) while the camera angle and visual embellishments persist, helping users maintain context and orientation while immediately seeing changes in gradients \textbf{(T7)}.
For example, a scientist could quickly switch between interpolations of taxonomic and geochemical attributes at various grid sizes to identify relationships between them and find locations with interesting gradients to sample in future dives \textbf{(T1, T2, T8)}.
Users can also clip through the voxels (Fig.~\ref{fig:interp_view}D) to see patterns in the data in the interior of the interpolated region.
We provide two clipping methods: (1) clipping individual voxels by their interpolation value outside of a range chosen by the user using a double-ended range slider; and (2) clipping by latitude/longitude/depth using a single-ended range slider.
In (2), cuts are taken only from the intersection of all three clipping planes, and we also ensure users can flip the clipping plane to view cuts from any direction.
Finally, if a user is interested in the exact parameter values at a specific latitude/longitude, they can hover over the object and select a vertical core to save (Fig.~\ref{fig:interp_view}E) \textbf{(T8)}.
The results are output in JSON as a list of parameter values at each 1 cm horizon, matching the exact format of a real core sample.
These results can then be further analyzed, e.g., in future dives by collecting new cores, deriving sample values, and comparing them with the interpolated values at the same latitude/longitude given by \textit{DeepSee} \textbf{(T9)}.

\subsection{Implementation}
\label{sec:implementation}
\textit{DeepSee} is an open-source\footnote{\textit{DeepSee} code: \url{https://github.com/orphanlab/DeepSee}} Vue.js Single-Page Application, which provides the UI framework and styling.
We plot cores on top of PNG map backgrounds in the \textit{Map View} as well as draw bar charts in the \textit{Core View} using D3.js \cite{Bostock:2011:D3}.
To compute interpolations, we used SciPy's ND piecewise linear barycentric interpolator \cite{SciPy} and a Python implementation of a 3D discretized approximation for Sibson's natural-neighbors interpolation \cite{Park:2006:NaturalNeighbors}.
Then, we used Three.js to create a traditional 3D render scene with global lighting in the \textit{Interpolation View}.
We also wanted \textit{DeepSee} to be accessible while doing fieldwork in remote locations.
Thus, we used Electron.js to create a portable web browser environment as a standalone desktop executable for both Windows and Mac operating systems that can run with or without access to the internet.

\section{Usage Scenarios}
\label{sec:usage}

\aptLtoX[graphic=no, type=html]{
    \begin{figure*}[!t]
      \centering
      \includegraphics[width=\linewidth]{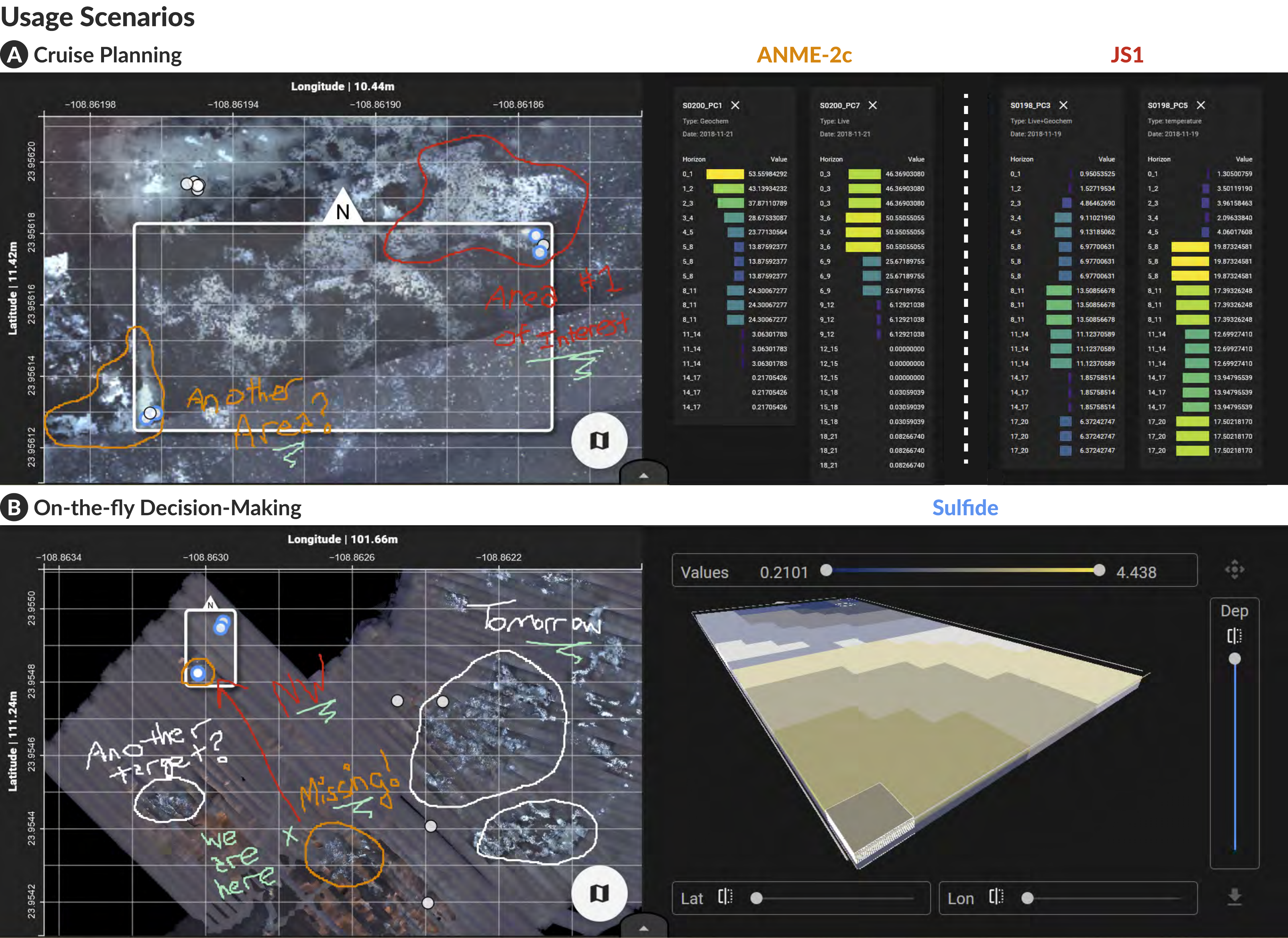}
      \caption{%
        Two usage scenarios showing how \textit{DeepSee} helps users make the most of limited sample data by supporting real-time annotation and interpolation in the tool, with data from Speth et al. \cite{Speth:2022:MicrobialAukaVents}.
        In \textbf{(A)}, our user maximized information between map and tabular data to determine where high-value cores containing both ANME-2c and JS1 are most likely to occur on future dives.
        In \textbf{(B)}, during an on-the-fly decision-making scenario where the seafloor changed over time, the \textit{Interpolation View} created a data-driven opportunity to correct course by interpolating unseen sulfide values and helping users find new targets.
      }%
      \Description{%
        Two images labeled (A) and (B) each showing the half-width double column main content area of DeepSee.
        Image (A) shows the Map View on the left and the Core View on the right.
        The Map View shows colored line annotations and the Core View shows ANME-2c and JS1 values for two core samples.
        Image (B) shows the Map View on the left and the Interpolation View on the right.
        The Map View shows colored line annotations and the Interpolation View shows a colored gradient of sulfide representing high values in the southwest corner transitioning to low values in the northeast corner.
      }%
      \label{fig:usage_scenarios}
    \end{figure*}
}{
}

In this section, we illustrate how \textit{DeepSee} can be used to study deep ocean microbial ecology via two example usage scenarios (Fig.~\ref{fig:usage_scenarios}).

\subsection{Scenario: Pre-Cruise Planning}
\label{sec:scenario_cruise_planning}
A graduate researcher is planning a follow-up cruise with their lab to collect cores at a site visited previously.
Based on prior sampling, they want to determine where cores with a high abundance of both ANME-2c archaea and JS1 bacteria might be (Fig.~\ref{fig:usage_scenarios}A).

They start by selecting all cores in the \textit{Map View} and ``ANME-2c'' in the \textit{Core View}, then examining the ANME-2c distribution down each core \textbf{(T4)}.
Using the \textit{Core View} instead of the filters in the \textit{Map View}, they locate four cores with high ANME-2c abundance \textbf{(T2, T8)}.
They select these cores in the \textit{Map View} \textbf{(T6)} and find them clustered in pairs of two.
Wondering about the interactions between ANME-2c and JS1, they select ``JS1'' in the \textit{Core View} and notice that the two cores to the North also have a high abundance of JS1 \textbf{(T1)}.
Using the menus of the \textit{Map View} to load a photomosaic map of seafloor imagery on the fly while maintaining the current viewport, they note that this region has a white microbial mat where the cores were taken \textbf{(T5)}.
They annotate the \textit{Map View} by drawing circles around the two cores to the North and write a short message: ``Area of interest \#1'' \textbf{(T3)}.
They then screenshot the image and place the image in a document for future reference \textbf{(T9)}.

The researcher decides that for the upcoming dive, retrieving cores from areas near these two cores with ANME-2c and JS1 are a high priority for their project, and focus should be on areas with a white microbial mat visible on the seafloor in this region.
\textit{DeepSee} enabled the researcher to maximize the value of limited samples by integrating several different data types (maps, tables, annotations) to make informed decisions about where to sample in the future.

\subsection{Scenario: On-the-Fly Decision-Making}
\label{sec:scenario_on_the_fly_decision_making}
A chief scientist (PI) is currently on a cruise, working with a team of submersible operators to visit their next sampling location.
Unfortunately, once on the seafloor, they discover the site no longer hosts the characteristic white and orange microbial mats and adjacent chemosynthetic clams observed a few years prior.
The chief scientist must quickly select another nearby seep sampling site with a high probability of similar microbial composition to collect sediment cores (Fig.~\ref{fig:usage_scenarios}B).

Using \textit{DeepSee}, the PI quickly loads a photomosaic map in the \textit{Map View} from the last cruise to this site \textbf{(T9)} and uses the \textit{Map View} to filter for the local cluster of previous cores collected here \textbf{(T6)}.
Looking at the spread of previous cores and the photomosaic map together, they realize the submersible is at the far end of the previously sampled area, which is $\approx50$ meters from an area that may still feature microbial activity \textbf{(T5, T7)}.
To test this hypothesis, they select all of the cores and choose ``Sulfide'' in the \textit{Core View}, identifying which cores have measured sulfide gradients which act as a proxy for active seeps \textbf{(T2, T10)}.
The PI eliminates cores without sulfide concentration data and identifies several cores that span the width of a previous sampling area \textbf{(T8, T9)}.
Switching to the \textit{Interpolation View}, they select a natural neighbor interpolation at $77$ cm resolution and a VSUP color palette \cite{Correll:2018:VSUP}.
The PI sees that there is a predicted sharp subsurface gradient towards higher sulfide concentrations heading northeast across the previous sampling area \textbf{(T11)}.
Switching the predicted attribute to microbial taxa identified at the site, they also observe a similar trend in the relative abundance of a methane-oxidizing archaeon, which are found in areas of methane seepage \textbf{(T1, T2, T11)}.
Throughout, they annotate the \textit{Map View} to support their decision-making, document the process, and share the results with the team \textbf{(T3, T8)}.

The PI decides that the submersible should navigate northwest to the sampled area and sample at, if not past, the location of core collections from the previous cruise located in the region with the highest sulfide concentrations.
Having \textit{DeepSee}'s ability to visualize unseen gradients quickly with limited data in the field can help researchers to make smarter, more data-driven decisions.

\section{Evaluation}
\label{sec:evaluation}

Our goal for evaluation was to validate the effectiveness of \textit{DeepSee}'s technical design and visualizations \cite{Sedlmair:2012:DesignStudyMethods} in terms of maximizing scientific return.
To do this, we first asked our deep ocean research collaborators (P$1-5$) which user tasks \textbf{(T)} raised in Sect.~\ref{sec:design_challenges} and Table~\ref{tab:challenges_tasks} they accomplished during an initial expedition deployment of \textit{DeepSee} aboard a research vessel in the Gulf of California (Sect.~\ref{sec:deployment}).
After deployment, we conducted semi-structured interviews with the same collaborators (P$1-5$) and collected feedback on how \textit{DeepSee} broadly affected research outcomes, workflows, and communication between team members (Sect.~\ref{sec:feedback}).
Based on the deployment and interviews, we then discuss limitations of the current implementation (Sect.~\ref{sec:limitations}) and finally conclude with lessons learned from our design process and the deployment that can be broadly applied to designing future visualization systems in other domains (Sect.~\ref{sec:lessons_learned}).

\subsection{Cruise Deployment}
\label{sec:deployment}

\begin{figure}[!t]
  \centering
  \includegraphics[width=\linewidth]{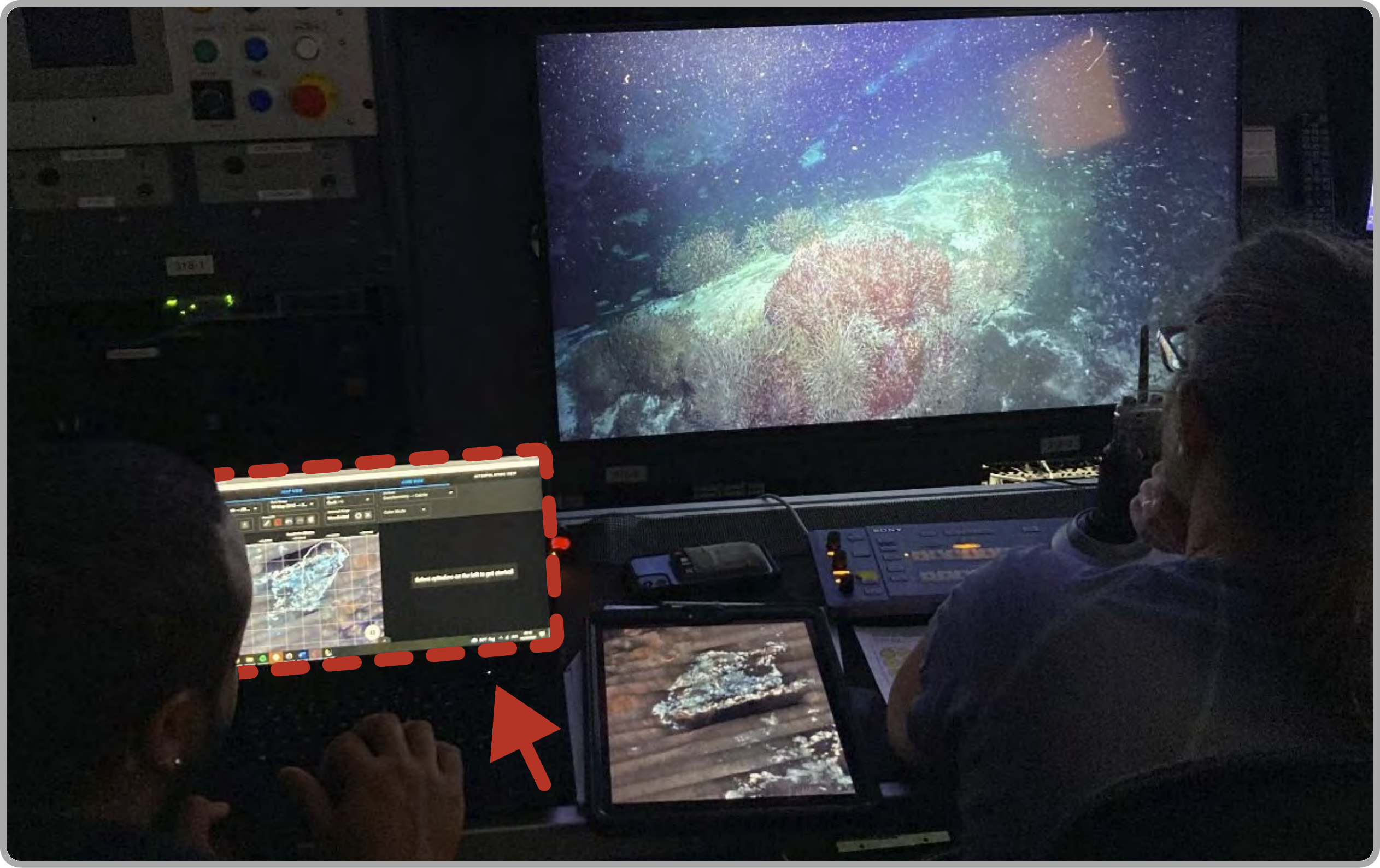}
  \caption{%
    \textit{DeepSee} being used shipboard during a cruise, helping the dive team locate where to sample cores around vents and seeps of the Pescadero Basin in the Gulf of California.
  }%
  \Description{%
    One photo showing two seated men in a dark, unlit room with digital monitors and submersible control equipment in front of them.
    One man is touching the keyboard of a laptop with the DeepSee interface showing on the screen.
    The other man is watching a large digital video monitor that shows a live color video feed from a submersible camera currently looking at the ocean floor.
  }%
  \label{fig:deployment}
\end{figure}

Following the evaluation methodology for design studies \cite{Sedlmair:2012:DesignStudyMethods}, \textit{DeepSee} was deployed on an expedition \cite{Parra:2021:DeepSeeBlogPost} aboard the research vessel R/V Falkor in the Gulf of California (Fig.~\ref{fig:deployment}) over the course of 15 days (Expedition FK210922) in 2021.
The goal was to help scientists characterize the geochemistry, biology, geology, and ecology of sediment-hosted hydrothermal vents and seeps of the South Pescadero Basin using mapping, ROV operations, and sample collection.
This area was first discovered in 2015 \cite{Paduan:2018:HydrothermalAlarconPescadero} and recently visited in 2018 on the R/V Falkor during expedition FK181031 \cite{Speth:2022:MicrobialAukaVents}.
Overall, DeepSee helped organize collected sample data from cruises into a central database for easy accessibility and comparison of various geochemical and microbiological community data.

Before embarking, core data collected in 2018 was processed and loaded into \textit{DeepSee} to visualize and contextualize each core's geochemical and biological attributes to one another and to their environment.
\textit{DeepSee} was used in several ways: (1) to familiarize newer scientists who were not aboard the 2018 cruise to the field site through the use of bathymetric maps and sample site locations \textbf{(T5, T6, T7)}; (2) to examine the sample data collected and characterize different sediment cores \textbf{(T3, T4)}; (3) to correlate different geochemical trends with microbial community data and to form hypotheses as to what was driving changes in microbial communities across the basin \textbf{(T1, T2)}; and (4) to identify scientific sites of interest to return to on the 2021 expedition to collect additional samples and examine how the conditions had changed over the course of three years \textbf{(T8, T9)}.
For example, the \textit{Interpolation View} supported gradient-based analysis \textbf{(T10, T11)}, as P$2$ described: \textit{``I used the Map View to see the spatialization of the cores, then the Interpolation View to see... if we sampled a core in a specific location, how would that change the certainty of the gradient estimation in that area? Where could we sample to add the most information?''}
P$3$ described a unique team-based use case of \textit{DeepSee} that emerged during this period:
\textit{``In the week before the Pescadero cruise... we all had DeepSee open together to look at sites and trends and to plan where to go. Seeing where the data is sparse, seeing where there is a lot of data, seeing map features without data points, seeing trends in geochemistry to follow out to some distance, looking for particular microorganisms in the sequence data by searching for species. It was quite helpful to have map data and sequence data in one place and see details on demand.''}
Overall, the scientists were able to test every task we designed \textit{DeepSee} for during pre-cruise planning.

During the cruise, the \textit{Map View} was used in the ship's control room as a visual reference by matching up the current dive location to data associated in any area that the ROV had dived previously in 2018 \textbf{(T5, T6, T7, T8, T9)}.
P$4$ described their typical workflow in the control room during a dive as well as having \textit{DeepSee} in parallel: \textit{``I was on the side as the data annotator for notes, screenshots, etc. I would pull up reference images and collect notes as the lead researcher would give orders and direct the group. I had DeepSee up and ready to go to answer questions. For example, if we returned to a previous area, we compared the data from before to what was there now. If it was a new area, there was not much to use DeepSee for.''} \textbf{(T3)}
P$4$ also mentioned that \textit{DeepSee}'s dark color scheme helped \textit{``keep light to a minimum in the [control room]. I didn't realize how important that would be until I was in there.''}
While many data exploration and annotation tasks were performed during the cruise, tasks related to analyzing data collected on the fly for decision making (\textbf{T1, T2, T4, T10, T11}) were not able to be performed.
We provide feedback from scientists on the additional support needed to perform in-situ analysis in Sect.~\ref{sec:limitations} and synthesize design guidance to address these needs in Sect.~\ref{sec:lessons_learned}.

\subsection{Expert Interviews}
\label{sec:feedback}
After the cruise, we conducted individual semi-structured interviews ($1-3$ hours long each) with five domain experts in deep ocean research (P$1-5$).
These experts were the same collaborators involved in the design process and carried out the initial deployment \cite{Parra:2021:DeepSeeBlogPost}.
Very few scientists have access to the deep ocean sample data we utilized, limiting the number of potential candidates able to properly evaluate \textit{DeepSee}.
However, in line with prior design studies \cite{Goodwin:2013:UCDEnergyVis, Mckenna:2015:UCDVizSec, Eirich:2022:IRVINE}, it is common for visualizations that tackle very specific problems to be validated by only a few users.

Our five collaborators were a team of graduate students, postdocs, and research scientists broadly interested in fieldwork at the intersection of geology, geochemistry, and biology in the deep ocean.
The multidisciplinary team has worked together over several years on logistics around planning and executing cruises, collecting and organizing prior and current cruise data, and analyzing samples obtained during cruises.
Two collaborators self-identified as female and three self-identified as male.
We organized the interview questions and domain experts' feedback around three broad questions we directly asked each collaborator.
We then conducted inductive thematic analysis \cite{Boyatzis:1998:ThematicAnalysis} of the feedback to identify emergent themes that were discussed amongst all authors.

\smallskip
\begin{itemize}
    \item[\textbf{(1)}] \textbf{Research} -- ``How did using \textit{DeepSee} support your fieldwork research goals?''
    \item[\textbf{(2)}] \textbf{Workflow} -- ``How did using \textit{DeepSee} improve the science return of samples?''
    \item[\textbf{(3)}] \textbf{Teamwork} -- ``How did using \textit{DeepSee} help you with your tasks in your role on the team?''
\end{itemize}

\medskip
\noindent\textbf{(1) In pre-cruise planning, fluid interaction and aggregation between multiple types and scales of data collected long-term can help researchers visually discover more insights.  }

\smallskip
We first explored how visualizations supported fieldwork-driven research.
In pre-cruise planning, several capabilities should be prioritized in system design, including data aggregation across multiple disparate sources, comparing multiple data types, fluidly interacting between data scales \cite{Elmqvist:2011:FluidInteraction}, and keeping humans in the analysis loop through data simulations.
These capabilities can enable serendipitous discoveries from analyzing data collected over a long period of time.
When asked who in the research community could benefit the most from \textit{DeepSee}, P$5$ responded: \textit{``It's really fieldwork driven. If someone has been cataloging a field site for a decade, like a marine site, I would absolutely recommend it.''}

To support these claims, our collaborators provided several examples of visualizations playing a pivotal role in supporting their research goals.
The team found directly comparing the \textit{Map View} with the \textit{Core View} or \textit{Interpolation View} helped them discover ``holes'' in their sampling.
For P$4$, this rich history of sampling helped them familiarize themselves with a new hydrothermal vent site: \textit{```Oh, these were hot cores, so this must be a hot area'... I can identify areas to sample before I [return to a field site].''}
These features introduced potential for making serendipitous discoveries in the future, as P$4$ described while demonstrating how they would use \textit{DeepSee}: \textit{``I didn't realize Sulfide was so high in this area! Why is that? ...Well, it looks like based on ANME distributions [in the Core View] ...these taxa are in this area and very abundant!''}
As a microbial ecology lab, comparing interactions between geochemical and biological phenomena at multiple scales is also vital.
P$3$ specifically felt that \textit{DeepSee} fit in a role that \textit{``enabled spatial thinking for the middle range of analysis''} between the microscopic and global level.
P$2$ described the value of \textit{DeepSee} in supporting human-in-the-loop sampling scenarios: \textit{``Many times we can't be [at a site] or make plans in time. Having a tool like this helps you come up with a plan backed by data and communicate these plans. We can overcome language barriers. Almost like a virtual reality simulation!''}

\medskip
\noindent\textbf{(2) When planning fieldwork expeditions, integrating and visualizing 2D/3D data together can increase the science return on limited samples.  }

\smallskip
In supporting the team's research goals, we also found that \textit{DeepSee} contributed to fieldwork research workflows in several unique ways.
We found that visualization of complex data during pre-cruise planning was the most useful to the team.
To support this, the ability to visualize data in a single interface helped experts orient themselves to the field site, perform gradient-based analysis across the area, and seamlessly explore trends between 2D and 3D.
This ultimately increased the long-term value or science return on using sample data to drive fieldwork research workflows in the future.
Beyond the visualizations, P$4$ also found that preparing data for testing in \textit{DeepSee} catalyzed useful data work for the long-term, such as \textit{``formatting and standardizing data collection, and planning on what columns we need.''}

P$5$ summarized an example \textit{DeepSee} workflow used in the lab, highlighting how questions involved overlaying multiple data types: \textit{``We first find a site that we're interested in, a vent for example, then zoom in to those samples and select them. Then, we'll go to the Core View and look at the geochemical and microbial taxa values for the cores. We'll filter out different core types... then look at the bar charts [in the Core View]. For example, we might see levels that are higher than normal of ANME, so we switch to a seafloor photomosaic map [in the Map View] and maybe we don't see any visible surface expression. This then leads us to ask what the geochemistry is in this area, so we switch to Sulfide [in the Core View]... Time is precious, so this [ability to toggle between data] helps with deciding what data we want to collect from this site again and which we might not need.''}
Comparing \textit{DeepSee} with existing map tools, P$1$ reported being more comfortable and efficient with ArcGIS, R and Python, while P$5$ reported they preferred using the \textit{Map View} to \textit{``quickly orient myself to where samples are and to test the area over where we wanted to interpolate.''}
P$4$, in charge of data preparation, summarized both the startup cost of \textit{DeepSee} and the payoff for their team in the long run around catalyzing data aggregation: \textit{``There was initially a higher amount of work to consolidate the data... Once we were past that main push, it was easier to use DeepSee... Now we can see patterns in the old data that were hidden in an Excel sheet. If I have a hypothesis and need to show the data to test that hypothesis, DeepSee makes it possible in just a few clicks.''}
Analogously, P$5$ described a new use case for \textit{DeepSee} \textit{``as a repository of information... new researchers that come into the lab can look at [previous] data we've collected from a site right away.''}

\medskip
\noindent\textbf{(3) Simple, intuitive and modular visualizations can help different fieldwork team members rapidly solve a diversity of specific, directed research tasks.  }

\smallskip
Finally, because we designed for a single team with a rich diversity of roles and responsibilities, we investigated how \textit{DeepSee} impacted communication and teamwork.
We found that the team dynamic can be characterized by many diverse goals with limited time and resources, where individual wants and needs illuminate unique use cases that make modular tools like \textit{DeepSee} valuable.
Having multiple visualizations with simple and intuitive controls helped our collaborators address specific tasks, such as filtering and selecting specific sites of interest, performing predictive analysis of potential data trends for deciding where to sample, and learning a single software instead of multiple tools.

While one of the lab's main research goals is to study deep ocean microbial ecology, every team member has unique questions they want to answer.
For example, P$2$ told us that \textit{``DeepSee lets me preselect sample sites in a meaningful way. I can design samples I would like based on the data we've gotten back.''}
P$2$ further found value in the predictive capabilities of the \textit{Interpolation View}: \textit{``If we think a gradient exists in a certain area, we can fill in the data and see that... DeepSee can maximize time by really targeting questions about exploration to hone in on specific data when there is a history of data.''}
For P$3$, a key improvement was in creating a single interface for visualizing data, \textit{``reducing the need to learn additional software such as ARCGIS.''}
During cruises, rapidly changing environments and unyielding time constraints can create high-pressure situations.
P$5$ described a particular cruise environment prior to the initial deployment of \textit{DeepSee}: \textit{``When we went to Pescadero years ago, the site was dynamic. The coordinates from the ROV are not very accurate. There's a lot of stress aligning previous bathymetry maps to the current location. You're looking at previous log books and looking at screens in the control room and talking to the pilots of the ROV.''}
Multiple team members agreed that simplicity and familiarity with \textit{DeepSee} were critical in such situations.

\subsection{Limitations}
\label{sec:limitations}
Several data wrangling and analysis limitations were expressed during the interviews.
A common issue discussed was prerequisite knowledge needed to format the data; e.g., P$5$ faced challenges \textit{``inputting DNA sequence information. Knowing how to input data and the centralization of that data is challenging. It's often that one person ends up inputting data.''}
While combining multiple data types in a single interface was a design goal, P$1$ lamented a reduction in dimensionality: \textit{``Certain data types might have different data descriptions and having multiple columns and reshaping tables is a challenge... Do I want to try to fit existing data into a certain tool like DeepSee?''}
P$1$ also wanted taxonomic data stratified hierarchically, a feature common to many biological analysis systems: \textit{``There could be very rare individual units that are technically different taxa. This creates very sparse data where abundance is very small for most of the columns. We needed a way to pick and find different subsets of taxa based on those hierarchies.''}
Finally, P$4$ wanted to see \textit{``a photo or snapshot of what a core actually looked like. An in-situ photo would allow a person to see what the sea floor looked like, what other cores were around it.''}

The \textit{Interpolation View} provided a contextualized way to view the data, as most other similar programs visualize gradients in 2D without spatial context between the sampled cores.
However, there was a lack of confidence in the accuracy of the interpolation methods, as natural neighbors and linear gradients are unlikely to reflect spatially heterogeneous in-situ environmental processes.
Similarly, distance to the nearest sample as a proxy for uncertainty was not accurate enough to use for in-situ decision making.
Still, P$2$ suggested potential applications of Value-Suppressing Uncertainty Palettes (VSUPs) \cite{Correll:2018:VSUP} in the future as the team develops more accurate uncertainty quantification: \textit{``We can see where the data is not strong and potentially sample to strengthen our predictions.''}
Specifically, the researchers noted that the modular nature of \textit{DeepSee} will make incorporating more sophisticated interpolation methods (e.g., a Gaussian process such as Kriging \cite{Oliver:1990:Kriging}) and uncertainty measures (e.g., variance in the posterior predictive distribution of the Gaussian process) in the future straightforward.

One deployment goal that was not validated during \textit{DeepSee}'s maiden expedition was whether data could be input in real time to visualize, as many of the relevant geochemical and biological parameters could not be measured while on the cruise, requiring further analysis on land.
P$3$ reflected on this challenge, saying: \textit{``We talked about using DeepSee on the fly during the research expedition to load data right away. So far, that's been impractical. We don't have geochem or bio data to put in DeepSee while at sea. We also haven't had a cruise to test whether these challenges were due to personnel on board or based on limitations with the DeepSee interface. Everyone was overextended on the expedition and no one had time to do additional data entry on the computer.''}
However, the team felt this helped them understand the time and energy allotment needed to use \textit{DeepSee} for real-time data integration and visualization.
P$3$ further discussed how they envision \textit{DeepSee} being used on future cruises: \textit{``In the future, we'll have a full team of scientists, and we'll also have a manned submersible taking cores... We'll have time to load the day's cores into DeepSee to help plan for the next day... The important objective is having more people to do the same number of tasks... I could see this becoming one of the evening tasks: take the day's coordinates and load them into DeepSee.''}
We synthesize design guidance for supporting real-time decision making shipboard in the next section.

\subsection{Lessons Learned}
\label{sec:lessons_learned}
Based on the initial deployment of \textit{DeepSee} (Sect.~\ref{sec:deployment}), user feedback (Sect.~\ref{sec:feedback}) and limitations (Sect.~\ref{sec:limitations}), we synthesized guiding principles for enhancing the design of future visualization systems that aim to support fieldwork-driven research.

\medskip
\noindent\textbf{Prioritize data integration as a user task.  }
Developing visualizations that support fieldwork-driven research allowed us to solve interesting constraints before and during expeditions, including limited computational resources and reconciling existing data collections with data collected during expeditions.
However, even though we prioritized predictive capabilities and uncertainty visualizations in \textit{DeepSee} to address these constraints, we unexpectedly discovered that not designing for data integration support as a user task limited the ability for experts to adaptably update hypotheses and make tactical decisions in the field.
It was difficult to track data being added on the fly, as P$5$ described the challenge they encountered during deployment of translating a mental decision-making process into using \textit{DeepSee}: \textit{``You might need to change your hypothesis and your methodology on the fly. These challenges often happen with the lab advisor in their head making big decisions... it was hard to incorporate DeepSee into the decision-making process, especially when it’s in the lab advisor’s head.''}
A mixed-initiative approach \cite{Horvitz:1999:MixedInitiativeUI} could alleviate time pressure during decision-making by bringing relevant data to the foreground based users' interaction history, bridging the gap in sensemaking between concurrent mental and physical analysis processes \cite{Pirolli:2005:Sensemaking} to determine and communicate what to sample.
Prioritizing data integration capabilities as a user task when designing visualization systems for fieldwork-driven research can ensure end-to-end support for tactical decisions when deploying tools out in the field.

\medskip
\noindent\textbf{Visualize physical data in context of the environment.  }
\textit{DeepSee} fostered new skills in communicating and conceptualizing complex phenomena using intuitive data visualization techniques.
For example, interpreting 3D data in the \textit{Interpolation View} was made easier by representing the data as it looks in real life, i.e., shaped as a cylinder, or with realistic proportions of height and area.
This allowed P$2$ to think through difficult research questions: \textit{``I think about the structure of the world I am exploring... There is a lot of data with a lot of dimensions. When we have discrete data points and samples but the environment is continuous, how do we represent phenomenon? For example, sampling the seafloor, or other planets, creates discrete measurements of continuous phenomenon.''}
Employing direct manipulation helped P$5$ understand the data more intuitively: \textit{``I like being able to physically click around and associate a map with samples. Before, I had to manually input [markers] with GIS software. DeepSee automatically visualizes samples [as they are added, so] now I can click on a sample and see details on demand.''}
P$2$ felt that showing the data as it looks in real life helped illuminate patterns in the data that intuitively showed \textit{``holes in sampling that we want to fill in''}.
The team also felt the visualization techniques in \textit{DeepSee} could extend to other disciplines, such as terrestrial field work integrating drone imagery or future autonomous sampling missions on other planets.
Designing data visualizations to mirror real-life counterparts can improve people's ability to communicate and understand complex scientific phenomena.

\medskip
\noindent\textbf{Combine data types in new ways to bridge analysis gaps.  }
For team members such as P$4$ doing data work, \textit{DeepSee} made centralizing and aligning data a priority: \textit{``DeepSee forced us to consolidate data from the same cruise and different cruises into the same format and in one location.''}
The desire to overlay multiple data types over different temporal ranges (i.e., region-, core-, and sample-level data in Sect.~\ref{sec:workflow_and_data}) catalyzed interest in answering new research questions that were previously time-consuming and/or difficult.
For example, by engaging fluid interaction between multiple levels of data aggregation in the \textit{Map View} and \textit{Core View}, \textit{DeepSee} enables users to make data-driven decisions based on complex geochemical and taxonomic data distributed at the region, core, and sample level over time (Sect.~\ref{sec:scenario_cruise_planning}).
This capability mirrors feedback from our expert interviews (Sect.~\ref{sec:feedback}) that a key feature of \textit{DeepSee} for P$3$ was enabling \textit{``spatial thinking for the middle range of analysis''} between the microscopic and global level.
Further, combining multiple data types in a single interface enabled deep ocean researchers to maximize the scientific value on limited sampling.
For example, P$5$ expressed benefits in their role as scientist of tracking hypotheses and data and their changes over time when filtering by cruise in the \textit{Map View}: \textit{``How do we know we're not reinventing the wheel? DeepSee has really enabled me to ask whether I'm contributing scientific work that fits in the data we have already collected...''}
\textit{DeepSee} shows the value of designing for and around data requirements in improving the scientific return and longevity of visualization tools.

\medskip
\noindent\textbf{Design interactive visualizations to aid mental modeling.  }
Interactive visualizations as a research tool can help scientists gain insight into their own workflows by seeing their problems through a different lens \cite{Conlen:2018:DesignVAOperations}.
For example, the \textit{Map View} enabled core sample data tables to be plotted ``live'' for building mental models of spatial ecological processes, rather than only as a ``static'' output for post-hoc modeling.
Because of this, during the cruise deployment, P$3$ saw increased awareness in colleagues of the potential for using visualization tools out in the field: \textit{``I see a desire from researchers to use more and different data products live in the field. Before... we didn't look at mapping data as much live, we didn't look at sequence data as much on the ship during the dive.''}
In this way, \textit{DeepSee} demonstrates potential as a generative tool to think with \cite{Hendrie:2022:CollabMethodSciVis}.
Manipulating physical data at different scales while continuously maintaining context with the environment between the \textit{Map View} and \textit{Interpolation View} aided researchers in their mental modeling of complex physical processes, such as how ecological processes might spread under the sea floor.
As a burgeoning PI, P$2$ is excited to use \textit{DeepSee} as \textit{``a resource for my students to make more informed critical decisions in this line of work. This tool fills in that gap by giving them a visualization of the field to help foster intuition.''}
They elaborated on the effects of using data visualizations to test ideas live: \textit{``I think using DeepSee influenced the way I teach... [having DeepSee] fundamentally changes the questions we can ask and the information that is available by changing the way we see the data.''}
\textit{DeepSee} exemplifies interactive visualizations as an opportunity for scientists to iteratively define and refine questions through ``making to know'' \cite{Hendrie:2022:CollabMethodSciVis, Zimmerman:2007:ResearchThroughDesign} and for designers to leverage the intuitive nature of visualization to demonstrate the art of possibility.

\section{Conclusions and Future Work}
\label{sec:conclusion}

In this work, we conducted a design study with a team of domain experts in geology, chemistry, and biology to develop multidimensional visualizations of seabed sediment cores for studying deep ocean microbial ecosystems.
We presented \textit{DeepSee}, an interactive workspace that enables scientists to explore their previous sediment sampling history and decide where new sediment samples may yield the greatest scientific return.
By visualizing samples in the context of the environment, bringing together multiple data types in a single interface, and providing capabilities to predict values in unsampled locations, \textit{DeepSee} offers a new approach to visualizing 3D point samples  in fieldwork-driven science.

In the future, we envision adding enhanced data analysis capabilities such as supporting multiple levels of phylogenetic hierarchy as well as aggregating and visualizing hierarchical data in the \textit{Core View}.
Additionally, we aim to integrate new and more sophisticated models for large-scale interpolation over several tens or hundreds of meters, based on new research enabled by \textit{DeepSee}.
The design elements introduced by \textit{DeepSee} could also extend beyond operator-driven deep ocean sampling into similar domains, including studying remote terrestrial ecosystems as well as interplanetary exploration missions.
For example, as autonomous sampling scenarios led by an AI system become possible, we could leverage insights into how human operators work with a system like \textit{DeepSee} to train future autonomous sampling systems.

\begin{acks}
Funding for this work was provided by grants from the National Science Foundation including the STC Center For Dark Energy Biosphere Investigations and NSF-OCE 2048666 (to V.J.O.).
D.R.U. is a National Science Foundation-Ocean Sciences Postdoctoral Research Fellow (2126631), R.L.W. is an NSF Graduate Research Fellow, and S.A.P. is supported through a grant from the NASA FINESST program (80NSSC22K1336). 
V.J.O. is a science fellow in the Canadian Institute for Advanced Science (CIFAR) in the Earth 4D program.

Data included in this study was collected during cruise FK181031 on the R/V Falkor supported by the Schmidt Ocean Institute. 
AUV data have been archived at the Marine Geoscience Data System (MGDS) in the data compilation titled \textit{PescaderoBasin\_MBARI}\footnote{\url{https://www.marine-geo.org/tools/search/entry.php?id=PescaderoBasin_MBARI}}.

The research was carried out at in part at the Jet Propulsion Laboratory, California Institute of Technology, under a contract with the National Aeronautics and Space Administration (80NM0018D0004).
The development of \textit{DeepSee} was enabled by Data to Discovery\footnote{\url{https://datavis.caltech.edu/}}, a data visualization, art and design research initiative based at NASA Jet Propulsion Laboratory, California Institute of Technology and ArtCenter College of Design; this support is gratefully acknowledged.
Finally, we would also like to thank Jasmine Otto for her valuable discussions about this research.

Any opinions, findings, and conclusions or recommendations expressed in this material are those of the author(s) and do not necessarily reflect the views of the National Science Foundation.
\end{acks}

\bibliographystyle{ACM-Reference-Format}
\bibliography{main}

\end{document}